\documentclass[useAMS,usenatbib]{mn2e}

\usepackage{graphicx}
\usepackage{textcomp}
\usepackage{amssymb}\usepackage{hyperref}
\usepackage{amsmath,alltt}

\usepackage{multirow}
\usepackage{rotating}
\usepackage{lscape}
\usepackage{times}
\usepackage{hyperref}
\usepackage{pdflscape}
\usepackage{supertabular}
\usepackage{xcolor}
\usepackage{ulem}

\newcommand{\Msun}{\mathrm{M}_{\odot}}

\title[NS/MSP binaries in the outskirts of star clusters]{The dominant mechanism(s) for populating the outskirts of star clusters with neutron star binaries}
\author[Leigh N. W. C., Ye C.S., Grondin, S.M., Fragione G., 
Webb J. J., Heinke C.]{Nathan W. C. Leigh$^{1,2}$\thanks{E-mail: nleigh@amnh.org (NL)}, Claire S. Ye$^{3}$, Steffani M. Grondin$^4$, Giacomo Fragione$^{5,6}$, \newauthor Jeremy J. Webb$^{4}$, Craig O. Heinke$^{7}$
\\
$^{1}$Departamento de Astronom\'a, Facultad de Ciencias F\'sicas y Matem\'aticas, Universidad de Concepci\'on, Concepci\'on, Chile \\
$^{2}$Department of Astrophysics, American Museum of Natural History, New York, NY 10024, USA\\
$^{3}$Canadian Institute for Theoretical Astrophysics, University of Toronto, 60 St. George Street, Toronto, Ontario M5S 3H8, Canada\\
$^{4}$David A. Dunlap Department of Astronomy and Astrophysics, University of Toronto, Toronto, Ontario M5S 3H4, Canada \\
$^{5}$Center for Interdisciplinary Exploration \& Research in Astrophysics (CIERA), Northwestern University, Evanston, IL 60208, USA\\
$^{6}$Department of Physics \& Astronomy, Northwestern University, Evanston, IL 60208, USA\\
$^{7}$Physics Department, CCIS 4-181, University of Alberta, Edmonton, AB, T6G 2E1, Canada}

\begin{document}

\pagerange{\pageref{firstpage}--\pageref{lastpage}} \pubyear{2011}

\maketitle

\label{firstpage}

\begin{abstract}
It has been argued that heavy binaries composed of neutron stars (NSs) and millisecond pulsars (MSPs) can end up in the outskirts of star clusters via an interaction with a massive black hole (BH) binary expelling them from the core.  We argue here, however, that this mechanism will rarely account for such observed objects.  Only for primary masses $\lesssim$ 100 M$_{\odot}$ and a narrow range of orbital separations should a BH-BH binary be both dynamically hard and produce a sufficiently low recoil velocity to retain the NS binary in the cluster.  Hence, BH binaries are in general likely to eject NSs from clusters.  We explore several alternative mechanisms that would cause NS/MSP binaries to be observed in the outskirts of their host clusters after a Hubble time.  The most likely mechanism is a three-body interaction involving the NS/MSP binary and a normal star.  We compare to Monte Carlo simulations of cluster evolution for the globular clusters NGC 6752 and 47 Tuc, and show that the models not only confirm that normal three-body interactions involving all stellar-mass objects are the dominant mechanism for putting NS/MSP binaries into the cluster outskirts, they also reproduce the observed NS/MSP binary radial distributions without needing to invoke the presence of a massive BH binary.  Higher central densities and an episode of core-collapse can broaden the radial distributions of NSs/MSPs and NS/MSP binaries due to three-body interactions, making these clusters more likely to host NSs in the cluster outskirts.
\end{abstract}


\section{Introduction} \label{intro}

Neutron star (NS) binaries located in the outskirts of star clusters have puzzled astronomers for many decades. The reason is that these objects 
are 
much heavier than the mean stellar mass in most old star clusters, in particular globular clusters (GCs), such that they are expected to segregate into the core on short timescales due to dynamical friction. And yet, such NS 
binaries have indeed been observed outside of the cluster core in many Galactic GCs (see Tables~\ref{tab:table1} and~\ref{tab:table2} for a full list).

For example, the core-collapsed GC NGC 6752 is known to have two millisecond pulsars (MSPs) located beyond the cluster's half-light radius. The first, dubbed PSR J1911-5958A or 
NGC 6752 A, is located about 3.3 half-light radii from the cluster centre and is the farthest MSP to have ever been observed from the cluster centre \citep{damico02}.  It 
contains a canonical MSP in a compact binary with 
helium white dwarf \citep{ferraro03,Bassa03}
of mass 0.20 M$_{\odot}$ \citep{Bassa06,Cocozza06,Corongiu23}.  The binary has a circular orbit and an orbital period of 0.86 days.  \citet{ferraro03} suggests that the MSP must have been the result of a dynamical interaction, but the MSP was recycled before the putative interaction occurred, based on the cooling age of the white dwarf (WD) \citep[e.g.][]{sigurdsson03}.  The second,  PSR J1911-6000C or 
NGC 6752 C, is located at about 1.4 half-light radii from the cluster centre and is an MSP similar to NGC 6752 A, but 
lacks a  companion \citep[e.g.][]{damico02}.

Other examples of MSPs located at or beyond the half-light radius in their host star cluster include: 
\footnote{See \href{http://www.naic.edu/~pfreire/GCpsr.html}{http://www.naic.edu/~pfreire/GCpsr.html}, a full catalogue of cluster pulsars.} 
J0024-7201X in NGC 104 (1.03 half-light radii), 
J1748-2446J in Terzan 5 ($\sim$ 1.3 half-light radii), J1748-2021C and J1748-2021D in NGC 6440 ($\sim$ 1.0 and 1.2 half-light radii), J1801-0857D in NGC 6517 ($\sim$ 2.4 half-light radii), M28 F, M13 B, M15 C, 
B1718-19A in NGC 6342, NGC 6624 K, M30 B, and XTE J1709-267 which may be associated with NGC 6293 (see \citet{jonker04}).  
For a more detailed summary, please refer to Table~\ref{tab:table1} and \ref{tab:table2} below.  For the purposes of this paper, we will consider any object 
at or 
beyond the cluster's half-light radius as being in the ``outskirts". 


\begin{table*}
    \caption{GCs with pulsars located around and beyond the half-light radius (i.e., in the cluster outskirts).  All data is taken from the pulsars in GCs catalog at \href{http://www.naic.edu/~pfreire/GCpsr.html}{http://www.naic.edu/~pfreire/GCpsr.html}, except the cluster mass, which is taken from \href{https://people.smp.uq.edu.au/HolgerBaumgardt/globular/}{https://people.smp.uq.edu.au/HolgerBaumgardt/globular/}.}
    \begin{tabular}{ |c|c|c|c|c|c|c|c|c|c| } 
    \hline \\
        Cluster ID & Mass & Core radius & Half-light radius & Distance & No. of pulsars & No. of pulsars & PCC?  &  No. of pulsars  \\
         & (in $10^5\,M_{\odot}$)  &   (in arcmin)  &   (in arcmin)   &   (in kpc)   &  & in binary & & beyond the half-light radius  \\
        \hline
NGC 104	& 8.53  &    0.36 &  3.17 &	  4.5 &	29 &  19  & no  &  1  \\
NGC 6205 & 4.84 & 0.62 & 1.69 & 7.1 & 6 & 4 & no & 1\\
NGC 6342 & 0.377 & 0.05  & 0.73   &  8.5 &  2  & 1  &   yes  &   1  \\  
Terzan 5 & 11 & 0.16	& 0.72  & 6.9  &	43	&  24 & no  &   1   \\
NGC 6440 & 5.7	& 0.14	& 0.48	 &   	8.5 &	8 &	4	&   no   &   2  \\
NGC 6517 & 2.2	& 0.06	&  0.5	  &      	10.6  &	17  &  $>$2	 &   no  &  1  \\
NGC 6624 & 1.03 &	0.06	& 0.82	   & 	7.9	&  11	&  2	&    yes   &   1   \\
NGC 6626 & 2.7 &  0.24  &  1.97  &  5.5  &   14   &   10   &   no   &   1   \\
NGC 6752 & 2.61 &	0.17	& 1.91	    &	4  &	9  &	1  &   yes  &  2  \\
NGC 7078 & 5.18 & 0.14 & 1.00 & 10.4 & 9 & 1 & yes & 1\\
NGC 7099 & 1.21 &	0.06  &	1.03   &		8.1	 &  2	&  2	&    yes  &   1   \\    
\hline
    \end{tabular}
    \\
    \label{tab:table1}
\end{table*}

\begin{table*}
    \caption{Properties of pulsars located around and beyond the half-light radius (i.e., in the cluster outskirts).  All data is taken from the pulsars in GCs catalog at http://www.naic.edu/~pfreire/GCpsr.html}
    \begin{tabular}{ |c|c|c|c|c| }
    \hline \\
        Pulsar Name
        & Offset & Companion mass & Eccentricity & Spin period \\
         & (in arcmin)  &   (in $M_{\odot}$)   &  & (in ms) \\
        \hline
NGC 104 X & 3.83  & 0.42 &  0.0000005 & 4.77152\\
NGC 6205 B & 1.626 & 0.186 & 0.000002 & 3.52807\\
NGC 6342 A & 2.3 & 0.13  &  $<$0.005   &  1004.04\\  
Terzan 5 J & 0.948 & 0.39	& 0.35 & 80.3379\\
NGC 6440 C & 0.48	& -	& -	& 6.22693\\
NGC 6440 D & 0.57	& 0.14 & 0.0 & 13.4958\\
NGC 6517 D & 1.202 & - & - & 4.22653	\\
NGC 6624 K & 1.43 & -	& - & 2.768\\
NGC 6626 F & 2.794 & - & - & 2.45115\\
NGC 6752 A & 6.39 & 0.22 & 0.00000082 & 3.26619\\
NGC 6752 C & 2.70 & -	& - & 5.27733\\
NGC 7078 C & 0.944 & 1.13 & 0.681386 & 	30.5293\\
NGC 7099 B & 1.2 & 1.31 &	0.87938 & 12.98983\\    
\hline
    \end{tabular}
    \\
    \label{tab:table2}
\end{table*}

 A popular mechanism often invoked in the literature to explain the presence of heavy NS binaries at large cluster-centric radii is interactions with massive black hole-black hole (BH-BH) binaries. For example, \citet{colpi02} and \citet{colpi03} 
 proposed that the presence of 
 NGC 6752 A 
 in the cluster's outskirts is most likely explained by an interaction with a massive BH-BH binary in the cluster core.  The authors use this observation to argue for the presence of an intermediate-mass BH (IMBH)-stellar-mass BH binary in the core, with a primary mass $\lesssim$ 100 M$_{\odot}$ and a low-mass secondary closer to 5 M$_{\odot}$.  \citet{colpi03} confirms that ejection velocities capable of delivering the NS binary to its currently observed location could be reached. 
 NGC 6752 C
 could also be explained by such an interaction, but \citet{colpi02} speculate that it might have been ejected to its current position due to an ionization event with a rare high-speed star.  

 In 
 \S~\ref{doublebh}
 we challenge 
 the idea 
 that massive BH-BH binaries are 
 the likely cause 
 of NS binaries observed in the halos of their host star clusters.  In \S~\ref{alternate}, we explore several alternative mechanisms that could allow the
 retention of NS binaries that are 
 not typically invoked in the literature. We also show that indeed more probable mechanisms exist, most notably a three-body interaction involving the NS binary and a normal cluster star. Finally, in \S~\ref{Simulations}, we compare our analysis to the results of Monte Carlo $N$-body simulations for star cluster evolution performed using the state-of-the-art \texttt{Cluster Monte Carlo} code. We discuss our results and conclude in \S~\ref{discussion}.

\section{An interaction with a BH-BH binary?} \label{doublebh}

In this section, we explore the possibility that a massive BH-BH binary or intermediate mass BH (IMBH)-BH binary is responsible for ejecting a given observed NS binary into the outskirts of its host star cluster. This mechanism has been adopted as the preferred mechanism to explain NGC 6752 A's location in the outskirts of NGC 6752 \citep{colpi02,colpi03}.

Consider an interaction in which two 3-100 M$_{\odot}$ BHs interact with a compact NS binary, ejecting it into the cluster halo \citep[e.g.][]{colpi02,colpi03}. First, we assume that the BH-BH binary is very close to the cluster centre (as expected from dynamical friction), such that the NS binary is ejected on a roughly radial orbit.  This means that the NS binary stalls at a distance of $R =$ 8 pc from the cluster centre, as observed for NGC 6752 A (for example). We assume a total mass $M$ for the cluster of 10$^6$ M$_{\odot}$ and that its density profile can be described by a Plummer sphere with a core radius $a = 1$ pc.  The NS binary has a total mass of 2.1 M$_{\odot}$, since we adopt a NS mass of 1.5 M$_{\odot}$ from an approximate mean of the distribution in \cite{capano20} and a WD or companion mass of 0.6 M$_{\odot}$ (see \cite{tremblay2016} for more information about the observed WD mass distribution), 
and is sufficiently compact that the timescale for it to interact directly with other stars at its current location exceeds several Gyrs.  
We note that there is some freedom in the choice of masses for the particles involved in the interaction, depending on the precise formation mechanism considered.  For example, consider a scenario in which a WD companion to the NS formed after the hypothetical dynamical interaction.  Then, the other star involved in the interaction, apart from the NS, would most likely be a typical MS star (with a mass that exceeds the present-day turn-off mass).  After the interaction, the MS star could evolve, leading to mass transfer or a common envelope event that eventually formed the final NS-WD binary.  Throughout this paper, our choices for the particle masses reflect the observed mass distributions wherever possible.

Then, for an isotropic non-rotating cluster, the timescale for the binary to return to the cluster centre on a roughly radial orbit
can be estimated approximately from the fallback time \citep{webb18}.  To order of magnitude, this is equivalent to the crossing time, or:
\begin{equation}
\label{eqn:cross}
\tau_{\rm cross} = \frac{R}{v_{\rm c}},
\end{equation}
where $v_{\rm c}$ is the circular velocity at radius $R$.  Since $v_{\rm c} = \sqrt{GM(< r)/r}$, we have:
\begin{equation}
\label{eqn:cross2}
\tau_{\rm cross} = \frac{3{\pi}}{4G\bar{\rho}},
\end{equation}
and $\bar{\rho} = M(< R)/(4{\pi}R^3/3)$ is the mean density inside $R$.  For our Plummer sphere:
\begin{equation}
\rho = \frac{3Ma^2}{4{\pi}}\frac{1}{(r^2 + a^2)^{5/2}},
\end{equation}
and
\begin{equation}
M(< r) = M\frac{r^3}{(r^2 + a^2)^{3/2}},
\end{equation}
hence
\begin{equation}
\bar{\rho} = \frac{3M}{4{\pi}(R^2 + a^2)^{3/2}}.
\end{equation}

Plugging in the required numbers into Equation~\ref{eqn:cross2}, we find a crossing time of $\sim$ 11,000 years.  For comparison, we can consider a more average cluster and adopt a cluster mass of 10$^5$ M$_{\odot}$, but this calculation yields a very similar crossing time of $\sim$ 34,000 years.  For very radial orbits, the NS binary will most likely be disrupted on its return to the cluster centre by another interaction with the central massive BH-BH binary, since both objects should return approximately to their point of ejection \citep[e.g.][]{leigh18}.  It is unlikely that the NS binary would be observed before returning to the cluster centre.  Thus, the probability of observing such a system is low, if the NS binary is ejected by the BH-BH binary when close to the cluster centre.  If the lifetime of the binary is of order the cluster age, the probability of actually observing it at any given time in this scenario is only $\sim 10^4/10^{10} \sim 10^{-6}$ assuming it takes a crossing time for the NS binary to return to the core.  We note that this assumes that the BH-BH binary gets a recoil decided by linear momentum conservation such that the NS binary re-encounters the BH-BH binary close to $r = 0$ 
upon its first pass through the core, on a timescale much shorter than the timescale for mass segregation to operate. 

\begin{figure}
\includegraphics[width=85mm]{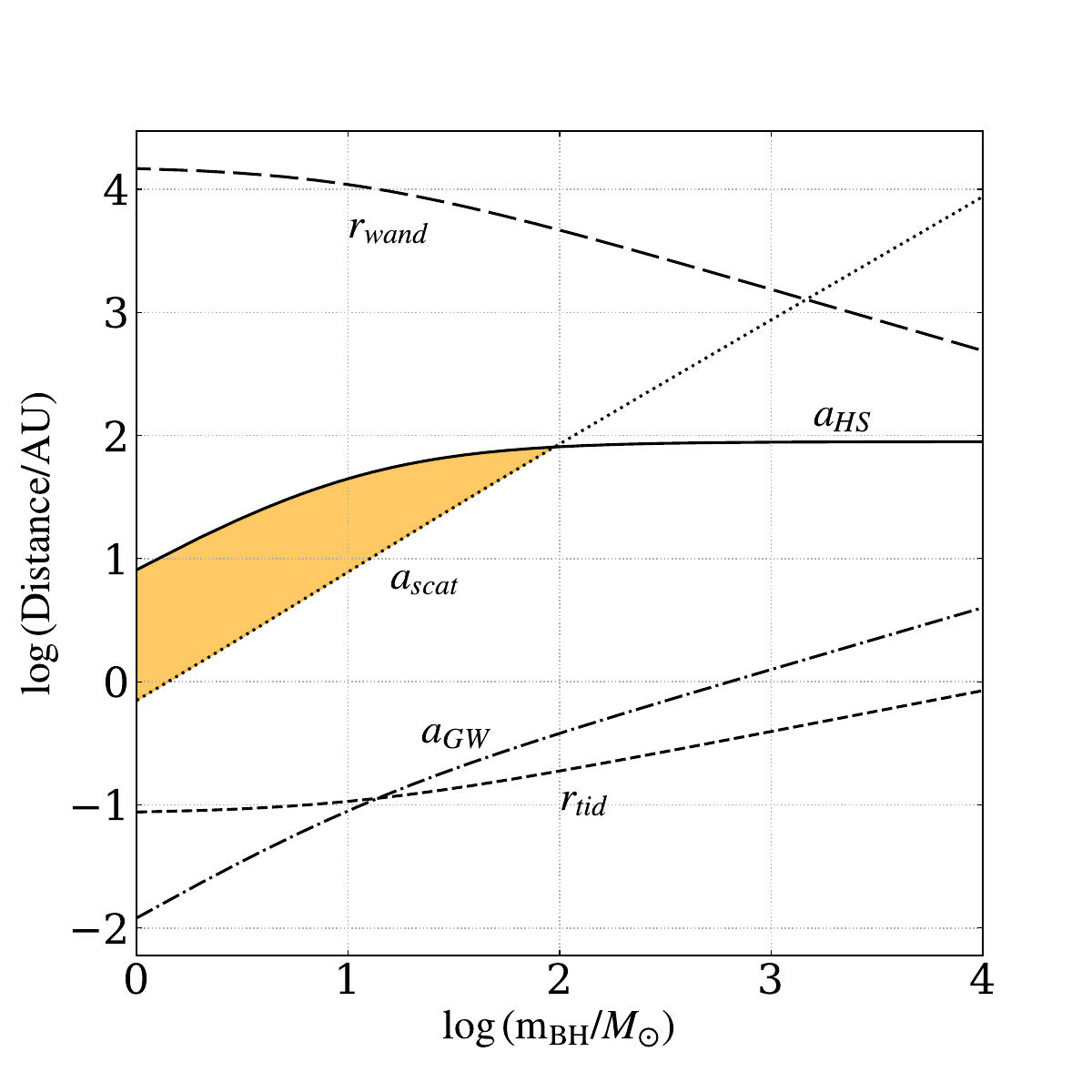}
\caption{We show as a function of the primary BH mass the critical orbital separation $a_{\rm GW}$ where the time for coalescence of the BH-BH binary due to gravitational wave radiation is equal to 10 Gyr, the critical orbital separation $a_{\rm scat}$ where the recoil velocity due to an interaction with the hypothetical NS binary is roughly equal to the escape speed from the core, the tidal disruption radius $r_{\rm tid}$ of the BH-BH binary (taken from \citet{colpi02}) and the wandering radius $r_{\rm wand}$ of the BH-BH binary due to Brownian motion (taken from equation 5.145b in \citet{merritt13} assuming a central density of 10$^5$ M$_{\odot}$ pc$^{-3}$ and a central velocity dispersion of 5 km/s).  We further show the hard-soft boundary of the BH-BH binary, or hardening radius, $a_{\rm HS}$, calculated using Equation 1 in \citet{leigh16}.  Note that all cluster and mass values are chosen to replicate the assumptions in \citet{colpi02} for the GC NGC 6752, except we adopt component masses of 0.6 and 1.5 M$_{\odot}$ for the NS binary while assuming a secondary BH mass of 3 M$_{\odot}$. 
Only for BH masses $\lesssim$ 100 M$_{\rm \odot}$ will the NS binary be retained in the cluster from a strong interaction, as indicated by the shaded area.  Here, the BH-BH binary orbital separation is most likely to be less than the hard-soft boundary $a_{\rm HS}$ but greater than the orbital separation corresponding to a recoil velocity equal to the core escape speed of 35 km s$^{-1}$ $a_{\rm scat}$ \citep{colpi03}, otherwise a higher ejection velocity is more likely.
\label{fig:fig1}}
\end{figure}

But what will happen when the recoiled NS binary is ejected by the BH-BH when it is away from $r = 0$?  Indeed, Figure~\ref{fig:fig1} shows that the wandering radius for an BH-BH binary tends to be of order $10^4$ AU, and this depends only weakly on the binary mass.  If originally radial then, without cluster rotation, we do not expect the NS binary's orbit throughout the cluster to deviate much from being radial \citep{webb19}.  However, it is possible that the NS binary has some finite orbital eccentricity less than unity (as is the case for a radial orbit) throughout the cluster, causing its return pass through the core to have an impact parameter that spares it from a direct interaction with the central BH-BH binary.  This could be the case either if the cluster has some rotation (since \citealt{webb19} showed that a kicked object will gain angular momentum from the cluster due to dynamical friction, giving rise to an eccentric orbit) or if the NS binary is kicked while off-center from $r = 0$.  

As estimated by \citet{colpi02}, 
the timescale for the binary to return to the core due to two-body relaxation is $\tau_{\rm df} \sim 7 \times 10^8 (1.6 M_{\odot}/m)$ years, which is again a small fraction of the total cluster lifetime but much longer than a crossing time.  For $m = 2.1$ M$_{\odot}$, the timescale is a little longer than 100 Myrs at the half-mass radius, yielding a probability of observing the system in the cluster outskirts at any given time of 10$^{8}$/10$^{10} \sim$ 0.01.  This is a lower limit, since we expect the timescale for dynamical friction to be even longer in the cluster outskirts relative to at the half-mass radius.  Once returned to the core, the NS binary should undergo a strong interaction with the central BH-BH binary on a timescale given 
by Equation 7 in \citet{leigh16}, or:
\begin{equation}
\label{eqn:si}
\tau_{\rm si} =  \frac{V_{\rm BH}}{\sqrt{3}\sigma_{\rm BH}\Sigma},
\end{equation}
where $V_{\rm BH}$ is the volume within which all stellar-mass BHs in the cluster are confined after mass segregation, $\sigma_{\rm BH}$ is the BH velocity dispersion and $\Sigma$ is the collisional cross-section.  This contributes of order $\lesssim$ 1 Gyr to the total timescale for a typical GC.  Hence, we estimate that, even if some eccentricity is imparted to the orbit of the NS binary, it should still most likely be destroyed by the central BH-BH binary well within the lifetime of the cluster.  This will occur on roughly a crossing time for interactions that occur very near the cluster centre of mass (i.e., without much wandering of the BH-BH binary) or some fraction of a core relaxation time near unity for off-centre interactions.  This is because if the BH-BH binary is at $r = 0$ then the NS binary will hit it upon returning to the core on its first pass through (due to conservation of linear momentum causing a recoil kick for the BH-BH binary in the opposite direction as the ejected NS binary), whereas off-centre collisions avoid this scenario, greatly prolonging the inspiral time of the NS binary.  We caution that the exact transition between these two scenarios is quite complicated and would require detailed N-body simulations to properly quantify, but for most of the relevant parameter space we expect the two-body relaxation time to be a better approximation since the NS binary is most likely to interact with the BH-BH binary away from the cluster centre.  

 We can quantify the above in another way.  Figure~\ref{fig:fig1} shows various critical distances pertinent to this problem as a function of the primary BH mass.  Specifically, we show as a function of the primary BH mass the wandering radius of the BH-BH binary, the hard-soft boundary for the BH-BH binary, the orbital separation corresponding to an inpsiral time of 10 Gyr due to gravitational wave radiation, the orbital separation yielding a recoil kick equal (from an interaction with the putative NS binary) to the cluster escape speed and the tidal disruption radius of the NS binary (see \citet{colpi02} and the figure inset for more details).  The orbital separation yielding a most probable recoil kick equal to the cluster escape speed is calculated using Equation 7.19 in \citet{valtonen06} by equating the cluster escape speed to the peak velocity of the centre of mass of the binary, or:
 \begin{equation}
 \label{eqn:peak}
  v_{peak} = \sqrt{\frac{2(M_t-m_e)}{5m_eM_t}}\sqrt{|E_0|},
 \end{equation}
 where $M_t$ is the total mass (i.e., the sum of the masses of all interacting particles), $m_{\rm e}$ is the mass of the ejected NS/MSP binary and $E_{\rm 0}$ is the total interaction energy.
 The key point is that the BH-BH binary orbital separation is most likely to be less than the wandering radius and, more importantly, the hard-soft boundary, but greater than the orbital separation corresponding to a recoil velocity equal to the core escape speed of 35 km s$^{-1}$ \citep{colpi03}; otherwise a higher ejection velocity for the NS/MSP binary is more probable. The BH-BH binary is dynamically hard and most likely to eject the NS binary to the outskirts (and not from the cluster) at less than the escape speed for primary BH masses $\lesssim$ 100 M$_{\odot}$ and only a narrow range of orbital separations, as shown in Figure~\ref{fig:fig1}.


Is the secondary mass in this scenario at all constrained?  In short, yes.  First, the secondary must be relatively massive such that an object heavier than the ejected binary should be left in orbit about the primary BH, otherwise the compact NS binary would be more likely to end up bound to it, ejecting the BH instead.  The mass of 
NGC 6752 A, for example, is sufficiently low that the secondary in the BH-BH binary need not be a BH.  A heavy neutron star could also get the job done roughly as easily, or perhaps even easier if NSs are more abundant in GCs than are stellar-mass BHs.  Second, in order for the NS binary to remain \textbf{intact}, 
\citet{wang18} showed that mass ratios near unity are preferred, and the dependence of the survival probability on the mass ratio is rather steep.

Given the arguments presented in this section, we conclude that a relatively improbable interaction with a BH-BH binary is needed to delicately place a NS binary into the outskirts of a star cluster and have it remain there for longer than a crossing time.  This is because only for primary masses $\lesssim$ 100 M$_{\odot}$ and a narrow range of orbital separations should the BH-BH binary be both dynamically hard and produce a sufficiently low recoil velocity to avoid completely ejecting the NS binary from the cluster.  

Provided other mechanisms to put NS binaries into the cluster outskirts could also be operating with a non-negligible probability, our results are consistent with a scenario in which clusters with 
high NS binary, and especially MSP, frequencies should most likely host the lowest mass BHs and the fewest BH binaries, as found in \citet{leigh16} and \citet{Ye+2019}, roughly independent of their observed radial distribution.  This is 
due to 
two reasons.  First, interactions with BH-BH binaries, especially more massive ones, are more likely to entirely eject NSs and MSPs from clusters than simply launch them into the cluster outskirts.  Second, in clusters with lots of BHs, the BHs act as a heat source for the NSs, preventing them from mass segregating into the centre \citep{Ye+2019}.  
The second reason should be the case in all but the most massive clusters ($\gtrsim$ 10$^6$ M$_{\odot}$) with the longest relaxation times, since here (especially in the outskirts) the relaxation times can exceed a Hubble time such that observing NS/MSP in the cluster outskirts should be independent of any dynamics happening in the core. 

\section{Alternative formation pathways} \label{alternate}

Although the BH-BH ejection scenario 
alone has been extensively discussed 
in the literature, other mechanisms also exist.  We will argue some of these are more likely to explain the origins of PSR A and other NS binaries floating beyond the half-mass radius of their host cluster.  In this section, we begin by listing the alternative possibilities to explain the presence of a NS binary in the outskirts of a dense star cluster, before exploring each in more quantitative detail.


The possible formation mechanisms for a NS binary in the cluster outskirts include:
\begin{itemize}
\item A primordial binary system born in the cluster outskirts.  
\item A three-body interaction with a normal cluster star (i.e., a WD or MS star).
\item A four-body interaction involving 
a binary composed of 
normal cluster stars. 
\item The disruption of a stable hierarchical triple due to the accretion-induced implosion of the tertiary companion.  
\item A natal kick partially imparted to the binary centre of mass due to accretion-induced collapse.
 Here, the natal NS gets a kick due to asymmetric mass loss in the detonating progenitor at the time of supernova explosion, which also imparts momentum to the binary centre of mass motion due to asymmetric mass loss from the binary system itself.  
\end{itemize}


Let us now consider each of the listed mechanisms in more detail.

\subsection{A primordial binary}\label{subsec:primordial}

Following \citet{colpi02}, the simplest possibility is that the binary is a primordial system born in the cluster outskirts.  This scenario is unlikely, however, given that the timescale for the binary to segregate back into the cluster core due to two-body relaxation is shorter than the cluster age in all but the most massive MW GCs.  This is because the binary is more massive than a typical single star, such that it will segregate back into the core on a relaxation time, or:
\begin{equation}
\tau_{r}(m) = \frac{<m>}{m}\tau_r(<m>),
\end{equation}
where $<m> \sim 0.5$ M$_{\odot}$ is the mean stellar mass for an old stellar population and:
\begin{equation}
\tau_r(<m>) = 1.7 \times 10^5 M^{1/2} \Big( \frac{r_h}{1 \rm{pc}} \Big)^{3/2} \Big( \frac{1 M_{\odot}}{<m>} \Big) {\rm years},
\end{equation}
where $r_h$ is the half-light radius.  Taking $r_h = 5$ pc for a typical Milky Way GC \citep{harris96} and setting $\tau_r(<m>) = 10$ Gyr, we find that only clusters with $M >$ 7 $\times$ 10$^{6}$ M$_{\odot}$ will have sufficiently long relaxation times for a primordial binary born in the cluster outskirts to still be located there today, having avoided segregating into the core due to two-body relaxation. Hence, in massive clusters like 47 Tuc, it would take the longest for NS binaries to segregate back into the core due to its larger mass and hence longer relaxation time, but the timescale is still much less than a Hubble time.

Importantly, for 
NGC 6752 A, the preceding argument suggests that a primordial origin is unlikely to be at the root of its unusual location far out in the outskirts of its host cluster.  The cluster is less massive than 7 $\times$ 10$^{6}$ M$_{\odot}$, suggesting that the NS binary would have had sufficient time to segregate into the core if it were born in the cluster outskirts.  Importantly, however, this estimate is obtained using the two-body relaxation time at the half-mass radius, whereas NGC 6752 A is located over 3 half-light radii from the cluster centre, where the relaxation time can be roughly an order of magnitude longer due to the much lower density. Hence, NGC 6752 A is somewhat of an unusual case due to its very large distance from the cluster centre, and a primordial origin cannot be entirely ruled out for this system given a purely two-body relaxation-based argument. 

An independent argument against many NS/MSP binaries located in the cluster outskirts having a primordial origin is as follows.
It is highly unlikely for MSPs to be formed in the outskirts of clusters, due to the low encounter rate in those outskirts. We know that MSPs are $\sim$100 times more frequent in globular clusters than in the Galactic field, as are low-mass X-ray binaries (LMXBs) \citep{Clark75}. 
For a simple estimate, take the mass of the Galaxy as $6\times10^{10}$ $\Msun$, and the mass of all globular clusters as $3.8\times10^7$ $\Msun$ \citep{Baumgardt+2018}. We use the estimate of 30,000 MSPs in the Galaxy from \citep{Lorimer13}, and estimate the number of MSPs in Galactic globular clusters by extrapolating the MSP numbers estimated by \citet{Zhao22} in 36 globular clusters (600-1500) to the remaining Galactic clusters by stellar encounter rate \citep{Bahramian13}, giving 1000-2500 MSPs in all globular clusters. Thus we confirm that globular clusters produce 50-130 times more MSPs per unit mass than the Galaxy. 

If we assume these halo MSPs are formed primordially, then their frequency should be similar to the field (actually, it should be substantially less, considering that the cluster escape velocity is small in the  outskirts, so neutron stars would escape even more easily than from clusters; \citealt{Pfahl02}.)  Let us assume the mass of the cluster outside the half-mass radius produces MSPs primordially, while the cluster inside the half-mass radius produces MSPs through dynamics. Then we should find (at least) 100-1000 times more MSPs inside the half-mass radius than outside the half-mass radius. (This assumes that no MSPs originally located outside the half-mass radius dynamically segregate inside the half-mass radius over time.)
Thus, we should find $<<$1\%  as many MSPs in the cluster outskirts as in their cores. But of the MSPs in the Freire catalog, we see 11 outside the cluster half-mass radius, and 130 within the half-mass radius; of order 10\% lie in the outskirts. This is far more than can be explained by primordial formation.

\subsection{A three-body interaction with a normal cluster star}

\subsubsection{Most Likely Ejection Velocities}

\begin{figure}
\includegraphics[width=85mm]{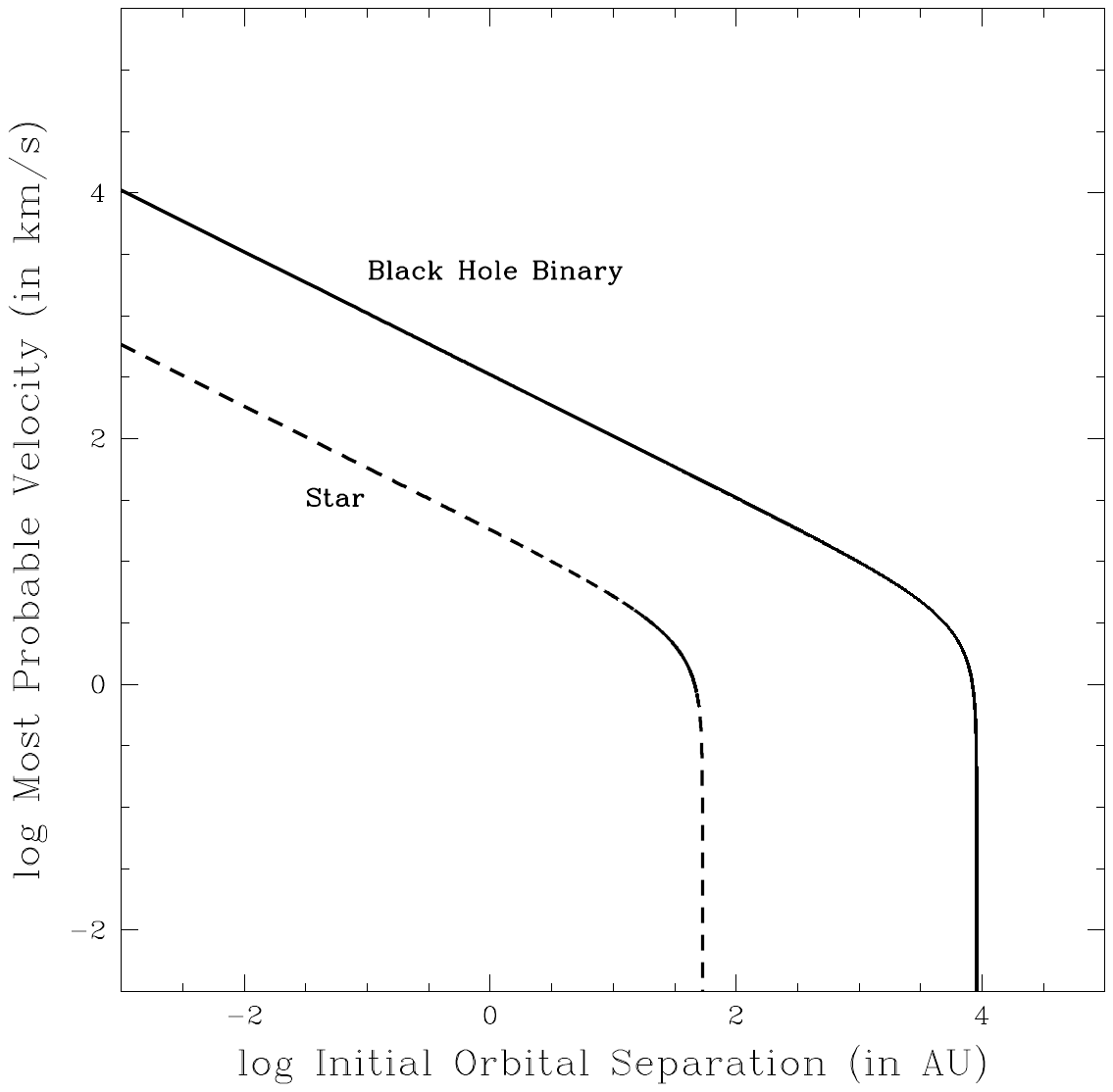}
\caption{We show the most probable ejection velocity as a function of the initial orbital separation of the ejected binary using Equation~\ref{eqn:peak}.  The solid black line shows the case where a BH-BH binary (composed of 100 and 5 M$_{\odot}$ BHs) ejects a compact NS binary (composed of a 1.5 M$_{\odot}$ NS and a 0.6 M$_{\odot}$ WD), whereas the dashed black line shows the same thing but for a three-body interaction involving the same NS binary and a 0.5 M$_{\odot}$ interloping single star. 
\label{fig:fig3}}
\end{figure}

Figure~\ref{fig:fig3} shows via the dashed line the most likely ejection velocity for a three-body interaction involving a NS (1.5 M$_{\odot}$), a WD (0.6 M$_{\odot}$; see \cite{tremblay2016}) and a normal MS star (0.5 M$_{\odot}$), leaving the NS and WD bound in a compact binary, as a function of the initial orbital separation (in this case, we assume the NS and WD were initially bound in a binary as well).
 This is done using Equation~\ref{eqn:peak} for the peak or most likely ejection velocity.  Note that we have corrected the final velocity for linear momentum conservation. For comparison, the solid line shows the same thing but for an BH-BH ejector.\footnote{We note that this is technically a four-body interaction, but it can be viewed as a three-body interaction if the NS binary is considered to be a single object due to its very compact orbit.}
As is clear, a more compact NS-WD binary is needed to achieve a higher ejection velocity than an BH-BH binary:  the BH-BH binary can achieve one order of magnitude higher velocities for the same initial orbital separation.  More importantly, a compact NS-WD binary can easily reach the cluster escape speed at an initial orbital separation of only 1 AU (and higher velocities are attainable for even more compact binaries). 

\subsubsection{Ejection Velocity Distributions}

To better compare the three-body interaction with a normal cluster star scenario to the historical scenario of a NS-WD binary interaction with a BH-BH binary, we make use of the \texttt{Corespray} particle spray code \citep{Grondin+2023}. Based on the theoretical three-body encounter framework presented in \cite{valtonen06}, \texttt{Corespray} samples the outcomes of three-body interactions within the cores of star clusters. Combining a cluster's orbital and structural parameters with a set of initial encounter configurations (e.g. system masses, orbital separations, binary binding energies, etc.), \texttt{Corespray}  ultimately produces statistical representations of the kinematics and positions of objects that have undergone three-body interactions in cluster cores \footnote{To learn more about the capabilities and installation of \texttt{Corespray}, refer to \cite{Grondin+2023} or visit \href{ https://github.com/webbjj/corespray}{ https://github.com/webbjj/corespray}.}. Using the previously discussed system in NGC 6752 as a template, we use orbital and structural conditions for NGC 6752 from \cite{Baumgardt+2018} and the same aforementioned encounter mass configurations. We then use \texttt{Corespray} to sample 50,000 three-body interactions for both cases over one azimuthal orbital period of NGC 6752 ($P_{orb}=132.162$ Myr), where the initial separations between interacting objects are randomly sampled between the semi-major axis of the binary and twice the mean separation of objects in the core (0.25 pc).

For the semi-major axis of the binary in Case 1, where three-body interactions are between a NS-WD binary and a normal MS star, we first randomly sample the binary's separation between the hard-soft boundary \citep{leigh22} and the contact boundary. The contact boundary is defined by assuming the same NS and WD masses as above, corresponding to a NS radius of $11$ km as approximated by \cite{capano20}, and a WD radius of $0.01 R_{\odot}$ approximated from \cite{provencal1998}. We then identify the range of initial separations that lead to the NS-WD binary hardening and having a final separation that is comparable to the observed separation of $a=0.025$ AU \citep{damico02} to within $10\%$. This range corresponds to separations between 0.02 and 0.05 au, which are randomly sampled to generate our final distribution of encounters.

For Case 2, the NS-WD binary is treated as a single object and it is the separation of the BH-BH binary that contributes most to the energy of the three-body system. When generating 3-body interactions, we consider two scenarios; (i) the BH-BH separation is equal to the hard-soft boundary, and (ii) the BH-BH separation is equal to half of the hard-soft boundary of the cluster. The hard-soft boundary for a BH-BH binary in a star cluster is given by Equation 1 in \cite{leigh16} and calculated to be $a=18.289$ AU, assuming masses of 100 $M_{\odot}$ and 5 $M_{\odot}$. 

\begin{figure}
\includegraphics[width=85mm]{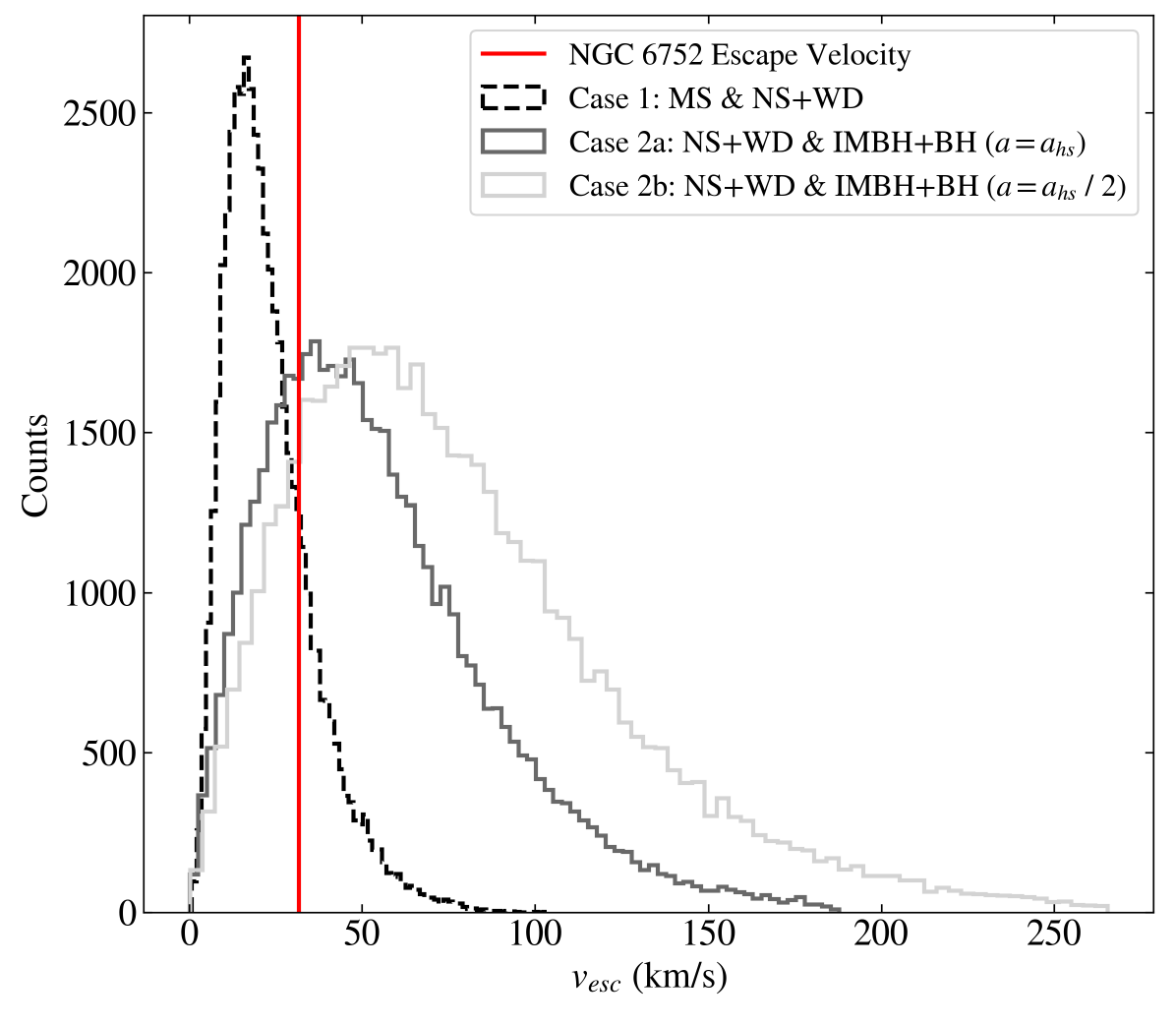}
\caption{We show the ejection velocity distributions for a three-body encounter composed of (1) a NS-WD binary interacting with a normal MS star (dashed line) and (2) a BH-BH binary interacting with a compact NS-WD binary (solid lines, with the the NS-WD binary being treated as a single compact object). In Case 2, we consider two additional sub-cases, where the binding energy is equal to the binding energy at (a) the hard-soft boundary and (b) half the hard-soft boundary in \citet{leigh22}. 
Both sets of $N=50,000$ three-body encounters are simulated using \texttt{Corespray} \citep{Grondin+2023}, 
where the \citet{Baumgardt+2018} escape velocity of $v_{esc} \sim 31$ km/s for NGC 6752 is indicated with a red line. It is clear that NS-WD binaries are more likely to be ejected from NGC 6752 when interacting with both soft and hard BH-BH binaries than with normal MS stars, providing evidence that off-centre NSs are more likely produced by the latter type of interaction.
\label{fig:vescdist}}
\end{figure}

Figure \ref{fig:vescdist} illustrates the NS binary ejection velocities from the \texttt{Corespray} simulation for the two cases \citep{Grondin+2023}. For the case of a NS-WD binary interacting with a normal cluster star, only $\sim 21\% $ of NS-WD binaries are given strong enough kicks that lead to them escaping the cluster. The probability of cluster escape for the case of a NS-WD binary interacting with a BH-BH binary is higher, where approximately $74\%$ and $86\%$ of all NS-WD binaries escape the cluster when the BH-BH binary's separation equals the hard-soft boundary and half of the hard-soft boundary \citep{leigh22}, respectively. Hence, we conclude that interactions with a BH-BH binary have higher chances of ejecting a NS-WD binary from a cluster than retaining it, for almost all ranges of BH-BH binary masses. Conversely, interactions with normal stars are more likely to result in a NS-WD binary remaining bound to the cluster, providing a possible explanation for the location of a NS binary in the outskirts of a GC.  In addition, using Equation A10 in \citet{leigh11}, such a single-binary interaction should occur roughly once every Myr, assuming a binary fraction of 10\%, a core radius of 1 pc, a core number density of 10$^6$ M$_{\odot}$ pc$^{-3}$, and a mean binary orbital separation of 0.05 AU (i.e., close to our assumed initial separation for the calculation corresponding to Case 1 above).  
For comparison, the analogous scenario involving a BH-BH binary should occur on a timescale closer to a Gyr, as given by Equation~\ref{eqn:si}.

\subsubsection{Post Interaction Behaviour}

After the interaction, the NS-WD binary will most likely sink back into the core on a relaxation time. This is the case even if the WD forms after the interaction (i.e., the secondary expands to become a giant and then a WD post-interaction), since for an initially compact binary, we do not expect the subsequent binary evolution to cause the final binary to widen significantly if at all (e.g., if a common envelope phase occurs this should most likely tighten the binary further). 

%

\subsection{A four-body interaction involving normal cluster stars}

In this scenario, a NS in a binary with a normal star (i.e., either a MS star or WD) undergoes an interaction with two other normal cluster stars of typical mass in a binary.  This scenario is unlikely to end in the production of two binaries \citep{leigh16} when four objects (i.e., two binaries) of similar mass and size interact.  Instead, the most likely scenario is that the two least massive objects are ejected sequentially as single stars, leaving the two most massive objects remaining bound in a binary \citep{leigh16b}.  Hence, since the direction of ejection should be more or less random and the velocity distribution for each single is the same as for a simple three-body disruption \citep{leigh16b}, this mechanism is roughly as likely to eject an NS binary as the analogous three-body interaction scenario with normal stars already considered in the previous section. 
For reference, using Equation A8 in \citet{leigh11}, such a binary-binary interaction should occur roughly every few tens of thousand years, assuming cluster parameters typical of massive GCs (such as NGC 6752 and especially 47 Tuc) namely a binary fraction of 10\%, a core radius of 1 pc, a core number density of 10$^6$ M$_{\odot}$ pc$^{-3}$, and a binary with a separation of 1 AU.  We expect the rate of single-binary interactions to dominate over binary-binary interactions in clusters with high central densities, however, since here the binary fraction tends to be $\lesssim$ 10\% \citep{leigh11,sollima08,leigh22}.  Hence, the analogous three-body interaction scenario is more likely.

With that said, however, \citet{leigh16b} showed that binary-binary interactions involving one wide binary and one compact binary tend to act as single-binary exchange interactions, with the heavy compact binary being exchanged into the wide binary, and ejecting one of its original binary companions in the process.  \citet{leigh16b} showed that, for reasonable initial assumptions, the recoil from the ejected single will often impart a kick of order a few tens of km s$^{-1}$ to the inner binary, perhaps enough for it to escape from the cluster core and into the outskirts.  Hence, this mechanism is likely to produce a velocity close to the required velocity to put the putative triple into the outskirts.  This scenario predicts that the compact NS binary should have a stable tertiary companion, though this seems to be ruled out in the case of NGC 6752 A by current pulsar timing \citep{Corongiu23}.

\subsection{The implosion of the tertiary of a hierarchical triple}

Consider a stable hierarchical triple star system containing a NS in the inner binary.  Hence, the system is composed of a NS binary in a compact orbit, with a third object orbiting it on a wide, stable orbit.  Let us assume that the outer tertiary is a white dwarf, and that the secondary in the inner binary is overflowing its Roche lobe, transferring mass to not only its NS companion but also the outer tertiary (this could occur if, for example, there is a common envelope event in the inner binary; see below).  If the tertiary is able to accrete, it could detonate as a supernova leaving behind no remnant (see \citet{leigh20} for more details).
If this happens, the remaining compact NS binary, formerly the inner binary of the triple, will be launched at the instantaneous orbital velocity.  Provided $v_{\rm orb} \lesssim v_{\rm esc}$, which should be the case for most stable triples given the need for a wide outer orbit in order to maintain dynamical stability, this could deliver a compact NS binary to the cluster outskirts.  

\begin{figure}
\includegraphics[width=85mm]{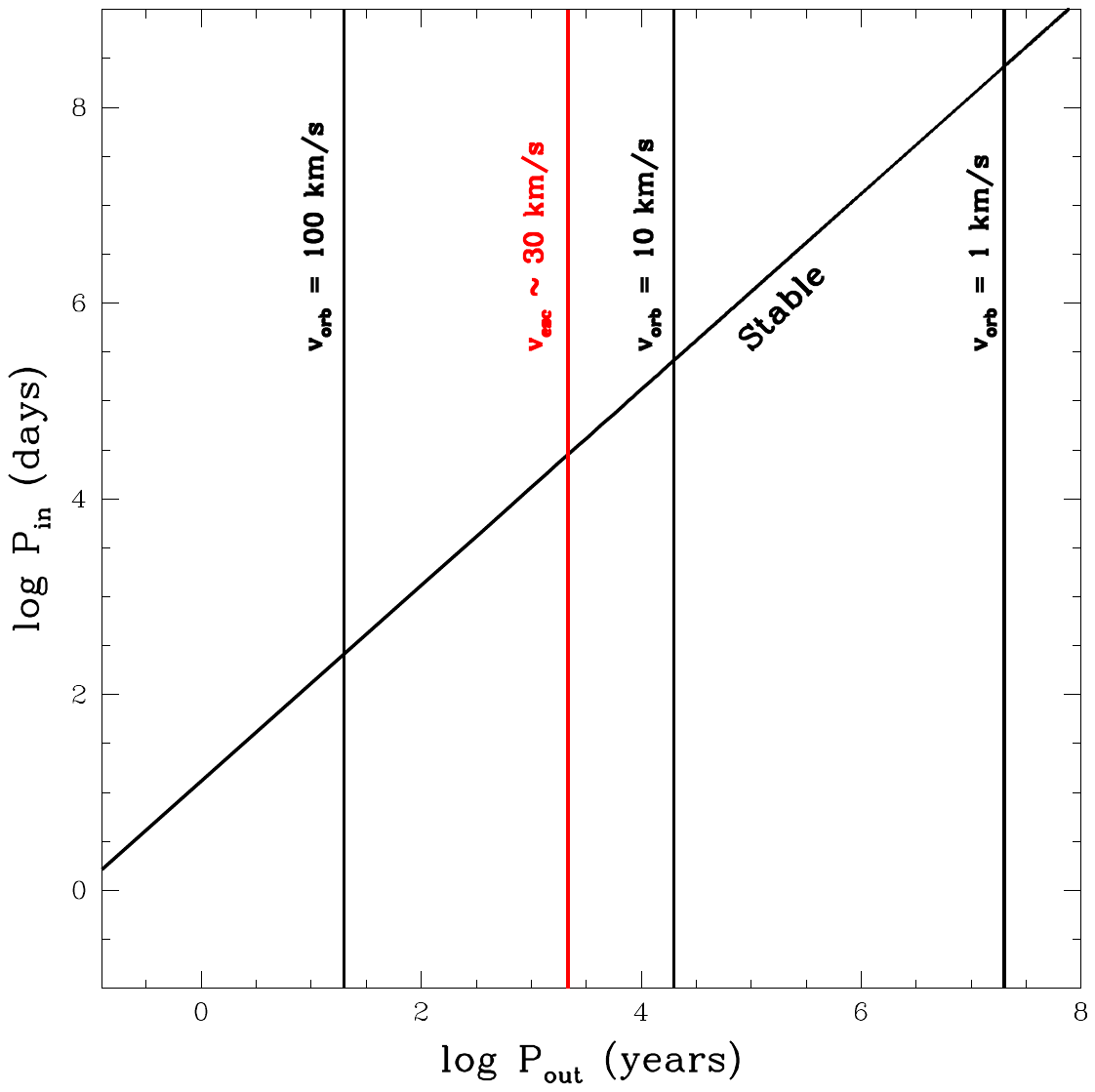}
\caption{We show with the black vertical lines in the P$_{\rm in}$-P$_{\rm out}$-plane the critical outer orbital period needed to achieve the indicated ejection velocity of the NS binary in the context of the disrupted triple scenario for getting NS binaries into the outskirts.  The red line shows the critical period for reaching the escape velocity; any objects falling to the left of this line will be ejected from the cluster.  
We assume component masses of 1.5 M$_{\odot}$ and 0.5 M$_{\odot}$ for the inner binary and a mass of 1.4 M$_{\odot}$ for the outer tertiary.  We further assume circular orbits.  Finally, the diagonal black line shows the boundary for dynamical stability (i.e., tertiaries are stable if they fall below the line) using the criteria from \citet{tokovinin18}.
\label{fig:fig2}}
\end{figure}

In Figure~\ref{fig:fig2}, we show the critical tertiary orbital period (i.e., the orbital period needed to achieve the indicated ejection/orbital velocities) for several different values of the ejection (i.e., orbital) velocity.  For this exercise, we assume a mass of 2.1 M$_{\odot}$ for the NS binary (as before) and a final mass of 1.4 M$_{\odot}$ for the outer tertiary just prior to detonation.  Assuming ejection velocities of 1, 10 and 100 km s$^{-1}$ gives critical tertiary orbital periods of $\sim$ 2.0 $\times$ 10$^7$, 2.0 $\times$ 10$^4$ and 20 years, respectively.  For comparison, assuming an average stellar mass of 0.5 M$_{\odot}$ and a velocity dispersion of 10 km s$^{-1}$, the critical orbital period is $\sim$ 5.9 $\times$ 10$^4$ years, which will be even longer in lower velocity dispersion environments. Hence, dynamically ``hard" triples can exist in clusters and still yield kick velocities with $v_{\rm orb} \lesssim v_{\rm esc}$, since $v_{\rm esc} \sim$ 10 - 50 km s$^{-1}$ for the densest and most massive Milky Way GCs.  Such wide triples should also be dynamically stable for any inner binary with an orbital period of hours or days \citep[e.g.][]{tokovinin18}.

Next, we wish to know how compact the outer orbit needs to be in order for the inner binary to undergo a common envelope (CE) event and transfer mass to the outer tertiary. 
To answer this, we compute the orbital velocity for the tertiary of our chosen triple.  Assuming an outer separation of 5 AU, the ejection velocity of the NS binary would be $\sim$ 31 km s$^{-1}$, or 22 km s$^{-1}$ for an outer separation of 10 AU.  Thus, if the tertiary accretes and detonates as a supernova explosion, leaving behind no remnant or causing it to disrupt dynamically, the compact NS binary could indeed be imparted with a kick of sufficient magnitude to deliver it to the cluster outskirts with $v_{\rm orb} \lesssim v_{\rm esc}$, provided the outer tertiary separation is in the range $\sim$ 5 - 10 AU.  This is evident from Figure~\ref{fig:fig2}, which shows that having a constraint on the period of the putative NS binary immediately constrains the period of the outer orbit of the hypothetical triple.  

With the above said, however, we caution that it is unlikely that the hypothetical inner binary will overfill its Roche lobe with a separation of order 10 AU, given that the turn-off mass in a typical GC is $\sim$ 0.8 M$_{\odot}$ or so.  This is likely the case even independent of any CE event, but detailed simulations would need to be performed to address just how much mass the putative tertiary might accrete.  We conclude that this mechanism is indeed a viable, but unlikely, option to put NS binaries into the outskirts of star clusters.  

A perhaps more likely albeit similar scenario could be invoked early on in the cluster lifetime if, instead of a WD, the outer tertiary is a massive star that ends its life as a supernova, causing the triple system to disrupt.  This does not leave much time, however, for the NS to be exchanged into the hypothetical triple system due to a dynamical channel and, as previously argued, the cluster dynamics must somehow be involved in the formation of NS/MSP binaries in order to explain their increased frequency in GCs relative to the field.


%

\subsection{A natal kick partially imparted to the binary centre of mass}

Consider a compact binary in a star cluster containing a primary WD and a secondary main-sequence star. As the secondary evolves, it will expand and transfer mass to the WD.  We assume that the primary ultimately explodes as a supernova, receiving a natal kick in the range of a few to several hundred km s$^{-1}$ due to asymmetric mass loss, and leaving behind a NS remnant.  If the kick direction opposes that of the orbital motion, then the binary can survive, and end up on a very compact eccentric orbit that should be rapidly circularized due to either tidal interactions or gravitational wave inspiral.  This scenario could also work in younger star clusters if a normal star is in a relatively compact binary with a massive star that explodes to produce a NS.

We imagine that some fraction $f$ of the expelled mass accelerates/decelerates the detonator directly, whereas the remaining mass fraction $(1 - f)$ acts to accelerate the binary centre of mass.  In order to properly distinguish between the two extremes (i.e., $f$ $\sim$ 1 or f $\sim$ 0), we would need to perform detailed hydrodynamics simulations and follow the mass-loaded expelled gas in detail.  To the best of our knowledge, such a study has yet to be done in the literature.

For now, let us take the simplest assumption, and set $f = 0.5$.  Then, if the mass is ejected in a direction that opposes the orbital motion, and a total mass $M_{\rm ej}$ is ejected, we can use conservation of linear momentum to compute the final velocity of not only the NS in its orbit but also that of the binary centre of mass motion.  Let us assume that the binary centre of mass is initially moving at 10 km s$^{-1}$ and expels in total 0.01 M$_{\odot}$ of gas at a speed of 100 km s$^{-1}$. 
Then by linear momentum conservation we have:
\begin{equation}
M_{\rm ej}v_{\rm ej} + (fM - M_{\rm ej})v_{\rm fin} = Mv_{\rm init},
\end{equation}
where $M = m_1 + m_2$ is the initial NS binary mass with $m_1 = 1.4$ M$_{\odot}$ and $m_2 = 0.8$ M$_{\odot}$.  This gives for the final ejection velocity of the NS binary:
\begin{equation}
v_{\rm fin} = (Mv_{\rm init} - M_{\rm ej}v_{\rm ej})/(fM - M_{\rm ej}) \sim 20 {\rm km s^{-1}},
\end{equation}
which is of sufficiently small magnitude to launch it into the cluster outskirts without ejecting it from the cluster.  


We conclude that the accretion-induced collapse of a WD primary in a binary could produce a sufficient recoil velocity to account for a compact NS binary observed in the outskirts of a star cluster.  How likely this mechanism is depends on the details of the supernova explosion and the probability of having a suitable progenitor binary in the cluster, both of which require further study to properly quantify.  The question of whether or not such an explosion can provide a sufficient kick to the binary centre of mass without unbinding it will be central moving forward.  We further caution that this mechanism also suffers from the same issue as discussed in the previous sections, namely that the cluster dynamics must be involved in the production of NS/MSP binaries in order to explain their much higher frequency in GCs relative to the field.  If this mechanism were operating with a substantial rate in GCs, than it should also do so in the field, over-producing the frequency of NS/MSP binaries in the field relative to what is observed.

\section{Simulations} \label{Simulations}

In this section, we present the results of Monte Carlo $N$-body simulations for GC evolution using the \texttt{Cluster Monte Carlo} code (\texttt{CMC}; \citealp[][and references therein]{CMC1}), which we use to assess whether or not GCs can eject NS binaries into the cluster outskirts.  We further use the models to assess the relative frequencies of the various mechanisms for putting NS/MSP binaries into the cluster outskirts discussed in the previous section.

\subsection{The code and initial conditions} 
\texttt{CMC} is based on the H\'{e}non-style orbit-averaged Monte Carlo method \citep{henon1971a,henon1971b}. It incorporates various relevant physics for cluster evolution, including two-body relaxation, strong dynamical interactions of singles and binaries, and tidal mass loss. Binary and stellar evolution is fully coupled to the dynamical evolution of the clusters and is calculated by the publicly available software \texttt{COSMIC} \citep{cosmic}, which is based on \texttt{SSE} \citep{sse} and \texttt{BSE} \citep{bse}. Strong three- and four-body gravitational encounters are directly integrated by the \texttt{Fewbody} package \citep{Fregeau+2004,Fregeau_Rasio_2007}, which includes post-Newtonian effects for BHs \citep{Antognini+2014,AS_Chen_2016,Rodriguez+2018a,Rodriguez+2018b}.

In particular, \texttt{CMC} simulates NSs and MSPs self-consistently following the treatments in \citet[][and references therein]{Ye+2019}, which showed good agreements with the spin periods and magnetic fields of observed pulsars. NSs are born in core-collapse supernovae (CCSNe), electron-capture supernovae (ECSNe), or accretion-induced collapses of WDs. \texttt{CMC} assumes that NSs born in CCSNe receive large natal kicks drawn from a Maxwellian distribution with a standard deviation $\sigma_{\rm{CCSN}}=265\,\rm{km\,s^{-1}}$ \citep{Hobbs+2005} due to asymmetries in the supernova explosion. On the other hand, NSs born in ECSNe or accretion-induced collapses receive small natal kicks drawn from a Maxwellian distribution with a standard deviation $\sigma_{\rm{ECSN}} = 20\,\rm{km\,s^{-1}}$ \citep{Kiel+2008}. All NSs are formed with spin periods and magnetic fields similar to the observed young radio pulsars. After their formation, NSs in binaries can be spun up to millisecond periods by angular momentum transfer during Roche lobe overflow \citep[][Eq.~54]{sse}, and their magnetic fields decay according to the `magnetic field burying' scenario \citep[e.g.,][]{Bhattacharya_vandenHeuvel_1991} where the magnetic fields decrease inversely proportional to the amount of mass accreted $(1+M_{acc}/10^{-6}\,M_{\odot})^{-1}$ \citep{Kiel+2008}. At the same time, isolated pulsars slow down through magnetic dipole radiation \citep{Kiel+2008}. For more details about the treatments of MSPs, see \citet{Ye+2019}.

As an example, we search for halo MSPs in \texttt{CMC} models of two clusters listed in Table~\ref{tab:table1}, NGC 6752 \citep[][their model 1a]{Ye+2023} which is a typical core-collapsed cluster \citep{harris96} and 47 Tuc which is a massive non-core-collapsed cluster \citep{harris96,Ye+2022}. These models closely match the respective clusters' observed surface brightness profiles and velocity dispersion profiles. The NGC 6752 simulation has an initial number of stars $N=8\times10^5$, virial radius $R_v=0.5$~pc, metallicity $Z=0.0002$, and Galactocentric distance $R_g=8$~kpc. Its stellar distribution follows a King profile with a concentration parameter $W_0=5$ \citep{King1966}. A standard Kroupa broken power-law \citep{Kroupa2001} between $0.08\,M_{\odot}$ and $150\,M_{\odot}$ is assumed for the initial mass function, and the model has an initial binary fraction of $5\%$. The 47 Tuc simulation has an initial number of stars $N=3\times10^6$, virial radius $R_v=4$~pc, metallicity $Z=0.0038$, and Galactocentric distance $R_g=7.4$~kpc. It initially follows an Elson profile \citep{Elson+1987} for stellar distribution with $\gamma=2.1$ where $\gamma$ is a free parameter of the Elson power-law slope \citep[][Eq.~8]{Ye+2022}. The simulation adopts a two-component power-law initial mass function with power-law slopes $\alpha_1=0.4$ and $\alpha_2=2.8$ for the lower- and higher-mass parts, respectively. Masses are sampled between $0.08\,\,M_{\odot}$ and $150\,\,M_{\odot}$ with a break mass at $0.8\,M_{\odot}$. The simulation assumes an initial $2.2\%$ binary fraction.


\subsection{General results}\label{subsec:sim_results}
In this section, we present the results of our CMC simulations for cluster evolution.  In particular, after discussing the radial distributions of NS/MSP binaries, we assess which of the mechanisms discussed in \S~\ref{alternate} are operating in the models and with what relative frequencies.

\begin{figure}
\includegraphics[width=85mm]{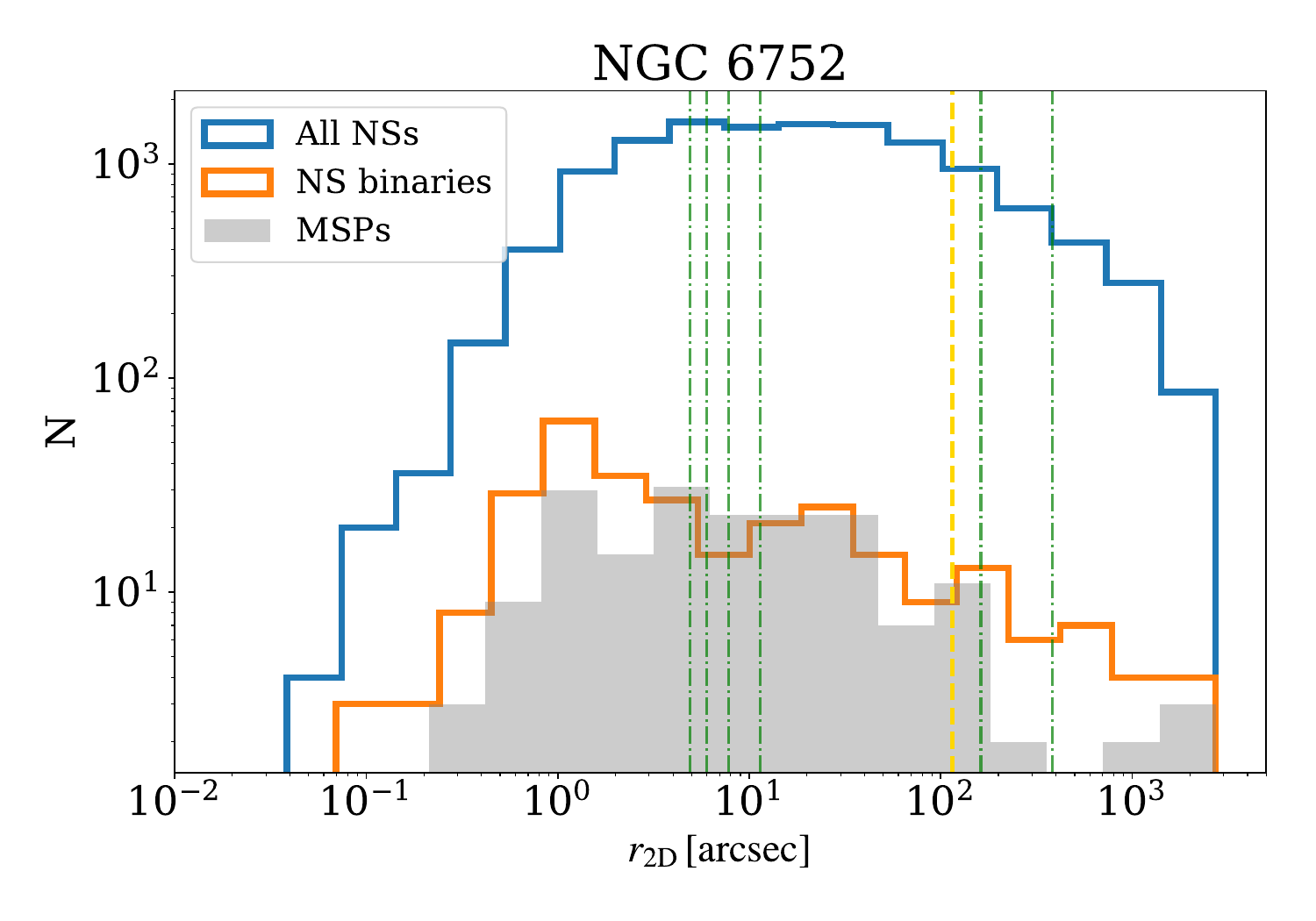}
\caption{Radial distributions of NSs (blue), NS binaries (orange), and MSPs (gray) from the NGC 6752 simulation. We combine the projected radial offsets from multiple time steps between 11 and 13.8~Gyr of the simulation for better statistics. The vertical green lines mark the offsets of the observed MSPs in NGC 6752 from \href{http://www.naic.edu/~pfreire/GCpsr.html}{http://www.naic.edu/$\sim$pfreire/GCpsr.html}. The vertical yellow line shows the observed half-light radius of the cluster \citep{harris96} 
\label{fig:radial6752}}
\end{figure}

\begin{figure}
\includegraphics[width=85mm]{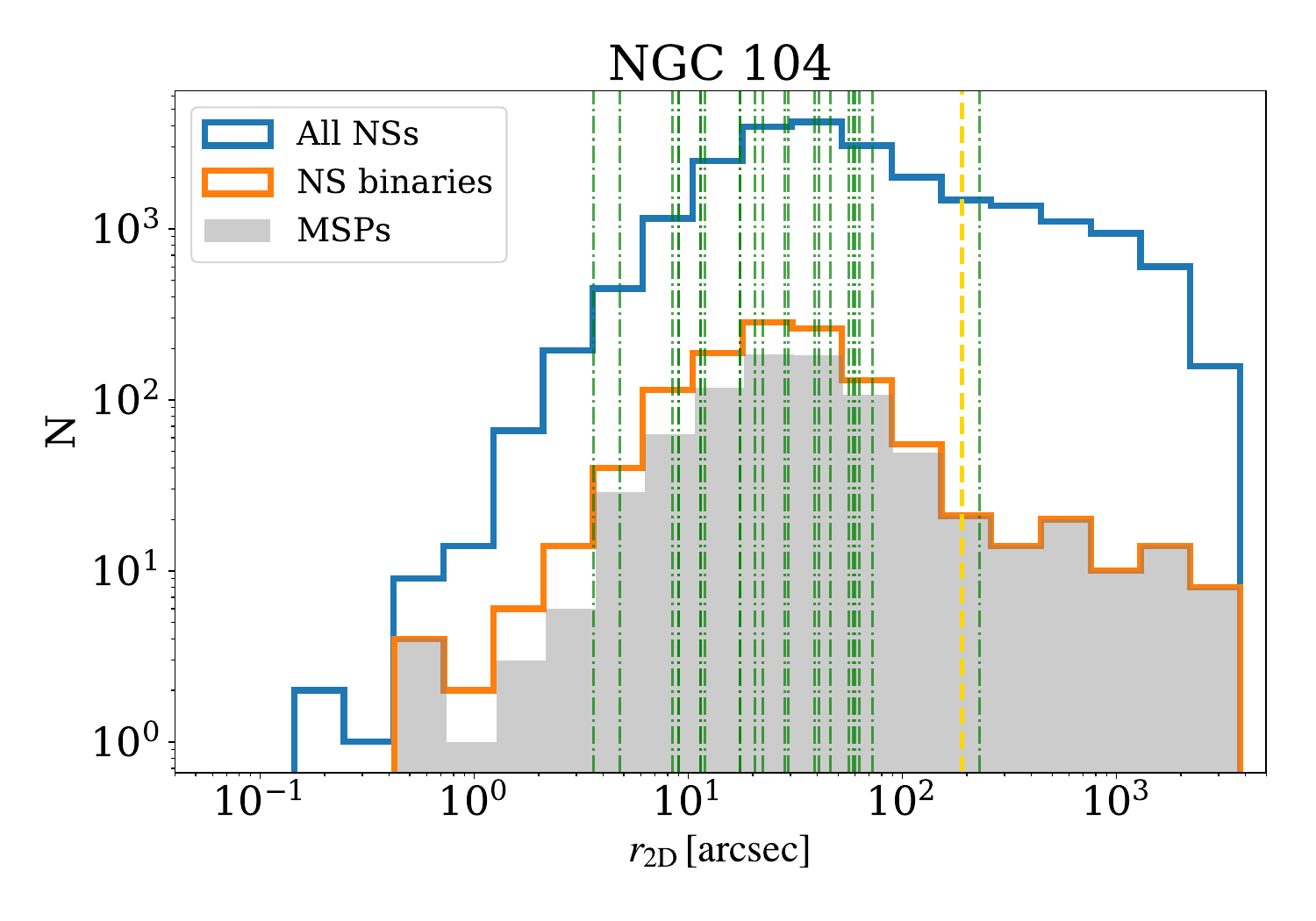}
\caption{Similar to Figure~\ref{fig:radial6752} but for radial distributions of NSs (blue), NS binaries (orange), and MSPs (gray) from the 47 Tuc simulation. We combine the projected radial offsets from multiple time steps between 9 and 12~Gyr of the simulation for better statistics. The vertical green lines mark the offsets of the observed MSPs in 47 Tuc from \href{http://www.naic.edu/~pfreire/GCpsr.html}{http://www.naic.edu/$\sim$pfreire/GCpsr.html}, and the vertical yellow line shows the observed half-light radius of the cluster \citep{harris96} 
\label{fig:radial47}}
\end{figure}

We show the projected radial offsets of NSs and MSPs from the NGC~6752 and 47~Tuc models in Figures~\ref{fig:radial6752} and \ref{fig:radial47}, respectively. Figure~\ref{fig:radial6752} includes all NSs and MSPs from 13 model snapshots (time steps) between 11 and 13.8~Gyr of the simulation for better statistics. The times roughly span the age observed for NGC 6752 \citep{Buonanno+1986,Gratton+1997,Gratton+2003,Correnti+2016,Souza+2020,Bedin+2023}. Overall, about six MSPs locate outside of the half-light radius of the cluster between 11 and 13.8~Gyr. At each of the 13 snapshots, we find between zero and three MSPs (all except one snapshot have at least one MSP), consistent with the observed number. These MSPs are ejected to the cluster halo directly through strong exchange encounters with WD binaries (either double WDs or WD-main-sequence star binaries), through natal kicks from the accretion-induced collapse of one of the components triggered by dynamical interactions with WD binaries (in this case it is an NS-WD binary), or through interactions with stellar-mass BH binaries or single stars such as main-sequence stars and WDs. Four halo MSPs are in binaries, where three binaries are in tight and circular orbits with very low-mass, WD-like companions ($\sim 0.01-0.03\,M_{\odot}$), and one is in an eccentric binary with a massive WD companion. In addition, most non-MSP NS binaries in Figure~\ref{fig:radial6752} are ejected to the outskirts by interactions with single WDs or main-sequence stars, with a few ejected by WD binaries.

These ejection mechanisms are consistent with those discussed in Section~\ref{alternate}. There is no IMBH in the NGC 6752 simulation, and there are $\sim 5$ stellar-mass BHs retained at the present day. It is also not surprising that MSPs can be relocated to the cluster halo through dynamical encounters with WD binaries. It has been shown that WDs dominate the cores of core-collapsed clusters \citep[][and references therein]{Kremer+2021wd} while most of the BHs formed in the clusters have been ejected through dynamical interactions. 


Similarly, Figure~\ref{fig:radial47} shows all NSs and MSPs from 17 snapshots between 9 and 12~Gyr of the 47 Tuc simulation, which is the age span predicted in \citet{Ye+2022}. A total of about six MSPs locate outside of the half-light radius of the cluster over this time scale, and at each snapshot, there are about three to five MSPs, 
consistent (within small-number statistics) with the one MSP seen outside the half-light radius. 
Different from NGC 6752, all of these MSPs are in tight and circular binaries with companion masses of $\sim 0.01\,M_{\odot}$. Note that the binary properties of the halo MSPs in the simulations are affected by the binary evolution prescriptions we adopt and may not match the observed properties exactly. Three of the six binaries are primordial and born far away from the cluster centre ($\gtrsim 4$~pc). The other three are formed through collisions with giant stars where the core of a giant star 
becomes 
a component star in the binary. The NSs in the latter binaries are formed in accretion-induced collapses (where a WD companion accretes from the core of the original giant star), and the natal kicks contribute to dislocating two of them from the cluster core (the third NS binary only appears very briefly at the outskirts, probably on an eccentric orbit in the cluster). On average over $\sim3$~Gyr, the fraction of MSPs from primordial binaries in the cluster outskirts is $\lesssim5\%$, 
somewhat larger than calculated in 
Section~\ref{subsec:primordial}. We also note that since 47~Tuc is more massive than most other GCs in the Milky Way \citep{harris96,Baumgardt+2018}, the number of halo MSPs formed in primordial binaries in other non-core-collapsed clusters 
will 
likely be closer to zero.

These binaries are not ejected to the cluster halo through recoil kicks from dynamical encounters as in NGC~6752, but rather because they are born in the outskirts (where the density is low and the relaxation time is long) and in part because the stellar-mass BHs retained in the cluster prevent the NSs from mass segregating to the cluster centre.  Unlike NGC~6752 which is core-collapsed and does not have many BHs retained, there are $\sim 200$ stellar-mass BHs retained in 47~Tuc at the present day, and there are no IMBHs \citep{Ye+2022}. Because of mass segregation, these BHs dominate the cluster core and act as energy sources through `BH binary burning', which supports the cluster from core-collapsing \citep[e.g.,][]{Kremer+2020bh}. At the same time, the lighter NSs are located further out and do not have many dynamical encounters \citep{Ye+2019}. Hence, in effect, the NSs and NS binaries are in the outskirts in the 47 Tuc model because (1) they are located far out in the outskirts where the stellar density is low and the relaxation time is long; and (2) the BHs heat the core, and this heat source in turn is transferred to the rest of the cluster, including the NSs, in part helping them to stay farther out in the cluster potential well for longer.  This effect is not included in the simple analytic estimates of the two-body relaxation time used in the previous sections.  This is different from what occurs in the NGC~6752 simulation, in which the NSs and NS binaries do segregate back into the core once the BHs have been ejected, but some of them can be ejected back out into the halo predominantly through three- or four-body interactions with normal stars and WDs. In this case, the absence of the BHs allows for core collapse to occur, which accelerates the rate at which NSs and NS binaries are ejected back into the cluster outskirts via single-binary interactions.

\begin{figure}
\includegraphics[width=85mm]{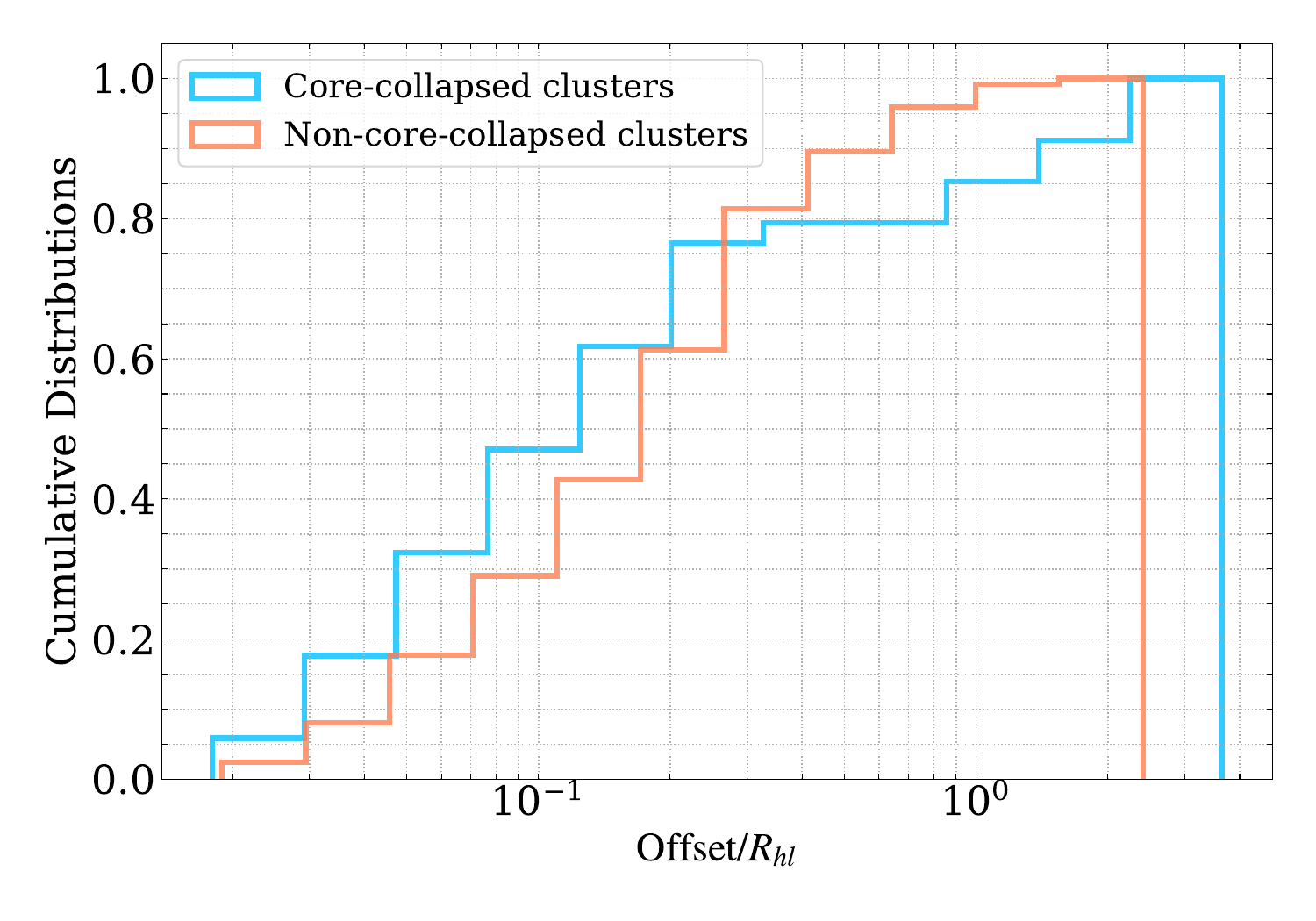}
\caption{The observed offset distributions in the unit of the host clusters' half-light radii of pulsars in GCs for both core-collapsed clusters (blue) and non-core-collapsed clusters (orange). Data taken from \href{http://www.naic.edu/~pfreire/GCpsr.html}{http://www.naic.edu/$\sim$pfreire/GCpsr.html}. We also estimate the offsets of pulsars in Omega Centauri \citep{Chen+2023omegacen} using the coordinates of the cluster center from \citet{harris96} and include them in the figure. 
\label{fig:obs_offset}}
\end{figure}

The aforementioned redistribution of NSs and NS binaries makes a prediction: the radial distribution of MSPs should be more extended in core-collapsed GCs relative to non-core-collapsed clusters. We can test this using the observed distributions of MSPs (see \href{http://www.naic.edu/~pfreire/GCpsr.html}{http://www.naic.edu/$\sim$pfreire/GCpsr.html}), and this is shown in Figure~\ref{fig:obs_offset}.  The radial distribution is slightly more extended for core-collapsed clusters \citep[see also, e.g.,][their Figure~2]{verbunt_freire_2014}, however, this result is not statistically significant. A KS test suggests that the two distributions may be drawn from the same underlying distribution, with a KS statistic of 0.16 and an associated p-value of 0.43. For comparison, an Anderson-Darling test suggests that the hypothesis that the two distributions are drawn from the same underlying distribution may be rejected at the 10$\%$ level. With that said, this comparison should be regarded carefully, since the prediction considered here does not account for other factors such as completeness, NS retention in clusters, etc.  These additional effects could be important and significantly affect our naive comparison.

\subsection{The dominant mechanism(s)}

Finally, we address which of the mechanisms discussed in \S~\ref{alternate} operate with the largest frequencies in our simulations, using the NGC~6752 simulation as an example. In general, single-binary interactions with MS stars and WDs tend to most commonly eject MSP binaries into the cluster outskirts, but binary-binary interactions also contribute (especially WD binaries).  This is no surprise since the timescale for single-binary interactions is shorter than that for binary-binary interactions for binary fractions $\lesssim$ 10\% \citep{leigh11}, and the binary fractions in our simulations are $\sim5\%$ for NGC 6752 and $\sim2\%$ for 47 Tuc.  We also find that natal kicks from the accretion-induced collapse of WDs contribute.  For example, of the six MSPs ejected to the outskirts in the NGC 6752 model, three of them are ejected by exchange encounters in binary-single interactions (one has a natal kick from accretion-induced collapse which may help get it further out), two of them experience a single-binary interaction as their last strong encounter and the last MSP is kicked to the outskirts via a natal kick from accretion-induced collapse and has a binary-binary interaction as its last strong encounter.


\section{Discussion and Summary} \label{discussion}

It has been argued in the literature that NS and MSP binaries can be ejected into the outskirts of star clusters via an interaction with a massive black hole binary that expels them from the core.  We challenge this idea in this paper and argue that this mechanism will only rarely account for such binaries.  
Only for primary masses $\lesssim$ 100 M$_{\odot}$ and a narrow range of orbital separations should a BH-BH binary be both dynamically hard and produce a sufficiently low kick velocity to retain the NS binary in the cluster.  
We explore several alternative mechanisms that would cause NS binaries to be retained in clusters, the most likely of which is a three-body interaction involving the NS/MSP binary and a normal star. 
 We expect normal stars (MS and WD) to be more common than BH-BH binaries, reducing the timescale for binary-single interactions with normal stars relative to that for interactions with BH-BH binaries.  We caution, however, that the precise answer will depend on the distributions of binary orbital properties (i.e., the orbital separation and mass ratio distributions), the frequency of BH-BH binaries and their mass distribution, and so on.

We argue in this paper that the NS binary NGC 6752 A, which lies far beyond the half-mass radius of its host cluster NGC 6752, was most likely placed there via a binary-single interaction with a normal MS or WD star.  This scenario is opposed to the system having been put there via an interaction with a massive BH-BH binary, as previously argued in the literature.  We naively expect for an old GC such as NGC 6752 that has experienced core-collapse that few BH-BH binaries should be left, with most having been ejected due to dynamical interactions with other BHs.  As argued previously, it follows that the timescale for binary-single interactions with normal MS or WD stars should be shorter than that for interactions with BH-BH binaries, even an IMBH-BH binary (see, for example, \citet{leigh14}).  All of these arguments indirectly suggest an inverse relationship between the frequency of BHs in clusters and that of NS/MSP binaries \citep[also see][]{Ye+2019}.  This is for two reasons. First, when lots of BHs are present, they provide a heat source to the cluster, delaying the NSs from mass segregating into the centre where they are more likely to undergo a dynamical interaction that exchanges them into binaries on a short timescale.  Second, if an interaction involving a BH and the NS/MSP or NS/MSP binary occurs, it is most likely to either exchange the BH into the binary or eject the NS/MSP binary from the cluster entirely.

The binary NGC 6752 A's overall properties match those expected from mass transfer from a subgiant, leaving a helium WD with the mass predicted by the \citet{TaurisSavonije99} relation \citep{Corongiu12,Corongiu23}. The low eccentricity suggests that the mass transfer occurred after the dynamical ejection event, although pulsar timing indicates that the WD spin is misaligned with the orbit, suggesting that the ejection happened after the mass transfer \citep{Corongiu23}.

We have utilized benchmark Monte Carlo $N$-body simulations of the clusters NGC 6752 and NGC 104 using the \texttt{CMC} code to test our results.  In NGC 6752, at about 12 Gyr, there are three simulated MSPs with offsets larger than the half-light radius, roughly matching the observed numbers.  Two of these MSPs are single and one is in a circular binary with a very low-mass WD-like companion (about 0.02 M$_{\rm \odot}$). The binary MSP is ejected to the halo from binary-single interactions with WD-WD binaries. 
We caution that, although the agreement between our simulated data and the observations is good for NGC 6752 at 12 Gyr, the number of MSPs in the outskirts is a 
sporadic 
function of time.  
However, for almost all timesteps between $\sim$ 11-13.8 Gyr, we find at least one halo MSP, suggesting that observing only a single MSP or NS/MSP binary is not altogether rare.

In the 47 Tuc simulation, we find that the NSs and NS binaries are in the outskirts at late times because the relaxation time can be long in the outskirts where the density is low and the BHs remain in the core where they act as a heat source.  This source of energy is ultimately transferred to the outer cluster regions, delaying a non-negligible fraction of NSs and NS binaries from segregating into the core.  It is important for the NSs to end up in the higher density core so that the timescale for them to be exchanged into binaries (and/or be dynamically hardened to a compact state) is sufficiently short.  This is different to what occurs in the NGC 6752 simulation.  Here, the NSs/MSPs and NS/MSP binaries have time to segregate into the core once the BHs have been ejected.  After this, some are ejected back out into the halo predominantly through three-body interactions with normal stars and WDs.  In the NGC 6752 case, the late-time absence of the BHs allows for core-collapse to occur, which accelerates the rate at which NSs/MSPs and NS/MSP binaries are ejected back into the halo.

Our simulations suggest that clusters that undergo core-collapse should experience a spike in the rate of single-binary interactions due to the increased central density \citep[see also, e.g.,][]{Ye+2019}.  This in turn increases the rate at which NSs/MSPs and NS/MSP binaries are ejected from the core due to dynamical interactions.  This implies that, for a given relaxation time (and hence total cluster mass and size), post-core collapse (PCC) GCs should be more likely to host MSPs and MSP binaries in the cluster outskirts.  This makes a prediction that can be tested observationally using the observed radial distributions of PCC and non-PCC clusters, namely that PCC clusters should show a broader radial distribution of NSs/MSPs during and after core-collapse.  
However, we find that MSPs in PCC clusters are observed to be only mildly more extended radially than are MSPs in non-PCC clusters.

We find from our simulations that the most common mechanisms to put NS/MSP binaries into the outskirts of GCs are single-binary interactions involving MS stars and WDs.  Binary-binary interactions also contribute but not as frequently as single-binary interactions since the timescale for single-binary interactions is shorter than that for binary-binary interactions for binary fractions $\lesssim$ 10\% \citep{leigh11}, and the binary fractions in our simulations are less than this.  We also find that natal kicks from the accretion-induced collapse of WDs contribute to putting NS/MSP binaries into the cluster outskirts.  For example, of the six MSPs ejected to the outskirts in the NGC 6752 model, five of them experience a single-binary interaction as their last strong encounter and the last MSP is kicked to the outskirts via a natal kick from accretion-induced collapse and its last strong encounter is a binary-binary interaction.

We can summarize our main conclusions as follows:

\begin{itemize}

\item In those clusters where the relaxation time is shorter than a Hubble time even in the outskirts, single-binary interactions involving MS stars or WDs are the dominant mechanism for putting NS/MSP binaries into the cluster outskirts.  This is supported both by the interaction energetics, which can give a sufficient recoil velocity kick to the NS/MSP binary centre of mass to put into the outskirts while also making it sufficiently compact to have an orbital period of order days or less (e.g., to produce NGC 6752 A).  

\item Interactions with BH-BH binaries are more likely to eject NS/MSP binaries from the cluster altogether based on energy-based arguments.  They can also operate on a much longer timescale than do normal single-binary interactions (i.e., involving only a NS and MS stars and/or WDs).

\item Natal kicks post-NS formation due to the accretion-induced collapse of a WD in a compact binary can also eject NS/MSP binaries from the core into the cluster outskirts, as found here both analytically and via \texttt{Cluster Monte Carlo} code simulations.

\item We find two reasons as to why some clusters might still be harbouring NS/MSP binaries in their outskirts after a Hubble time: 
 (1) Clusters with relaxation times in their outskirts that exceed a Hubble time could still host today NS/MSP binaries in their outskirts.  As argued in Section~\ref{subsec:primordial}, however, this primordial mechanism could only realistically explain a handful of NS/MSP binaries in the outskirts out of the total observed sample considered here (i.e., the Freire catalog).   (2) In clusters with lots of BHs, the BHs can act as a heat source in the core which feeds kinetic energy to the other stars and hence in part prolonging the NSs from mass segregating into the centre.

\end{itemize}



\section*{Acknowledgments}

NWCL gratefully acknowledges the generous support of a Fondecyt Regular grant 1230082, as well as support from Millenium Nucleus NCN19\_058 (TITANs) and funding via the BASAL Centro de Excelencia en Astrofisica y Tecnologias Afines (CATA) grant PFB-06/2007.  NWCL also thanks support from ANID BASAL project ACE210002 and ANID BASAL projects ACE210002 and FB210003. C.S.Y acknowledges support from the Natural Sciences and Engineering Research Council of Canada (NSERC) DIS-2022-568580. S.M.G. is partially supported from an Ontario Graduate Scholarship. G.F. acknowledges support from NASA Grant 80NSSC21K1722 and from NSF Grant~AST-1716762 at Northwestern University. CH is supported by NSERC Discovery Grant RGPIN-2016-04602. The authors also thank Maria Drout, who provided important neutron star insights that improved this study.

\section*{Data Availability}

The data underlying this article will be shared on reasonable request to the corresponding author.         


\bsp

\label{lastpage}


\begin{thebibliography}{99}

\bibitem[\protect\citeauthoryear{et al.}{2014}]{Antognini+2014} Antognini J.~M., Shappee B.~J., Thompson T.~A., Amaro-Seoane P., 2014, MNRAS, 439, 1079. doi:10.1093/mnras/stu039
\bibitem[\protect\citeauthoryear{Amaro-Seoane \& Chen}{2016}]{AS_Chen_2016} Amaro-Seoane P., Chen X., 2016, MNRAS, 458, 3075. doi:10.1093/mnras/stw503
\bibitem[\protect\citeauthoryear{Bahramian et al.}{2013}]{Bahramian13} Bahramian A., Heinke C. O., Sivakoff G. R., Gladstone J. C., 2013, ApJ, 766, 136. doi:10.1088/0004-637X/766/2/136
\bibitem[\protect\citeauthoryear{Bassa et al.}{2003}]{Bassa03} Bassa C. G., Verbunt F., van Kerkwijk M. H., Homer L., 2003, A\&A, 409, L31. doi:10.1051/0004-6361:20031339
\bibitem[\protect\citeauthoryear{Bassa et al.}{2006}]{Bassa06} Bassa C. G., van Kerkwijk M. H., Koester D., Verbunt F.,   2006, A\&A, 456, 295. doi:10.1051/0004-6361:20065181
\bibitem[\protect\citeauthoryear{Baumgardt et al.}{2018}]{Baumgardt+2018} Baumgardt H., and Hilker M., 2018, MNRAS, 478, 2. doi:10.1093/mnras/sty1057
\bibitem[\protect\citeauthoryear{Bedin et al.}{2023}]{Bedin+2023} Bedin L.~R., Salaris M., Anderson J., Scalco M., Nardiello D., Vesperini E., Richer H., et al., 2023, MNRAS, 518, 3722. doi:10.1093/mnras/stac3219
\bibitem[\protect\citeauthoryear{Bhattacharya \& van den Heuvel}{1991}]{Bhattacharya_vandenHeuvel_1991} Bhattacharya D., van den Heuvel E.~P.~J., 1991, PhR, 203, 1. doi:10.1016/0370-1573(91)90064-S
\bibitem[\protect\citeauthoryear{Binney \& Tremaine}{1987}]{binney87}
  Binney J., Tremaine S., 1987, Galactic Dynamics (Princeton:
  Princeton University Press)
\bibitem[\protect\citeauthoryear{Breivik et al.}{2020}]{cosmic} Breivik K., Coughlin S., Zevin M., Rodriguez C.~L., Kremer K., Ye C.~S., Andrews J.~J., et al., 2020, ApJ, 898, 71. doi:10.3847/1538-4357/ab9d85
\bibitem[\protect\citeauthoryear{Buonanno et al.}{1986}]{Buonanno+1986} Buonanno R., Caloi V., Castellani V., Corsi C., Fusi Pecci F., Gratton R., 1986, A\&AS, 66, 79
\bibitem[\protect\citeauthoryear{Capano et al.}{2020}]{capano20} Capano C. D., TewsvI., Brown S. M., Margalit B., De S., Kumar S., Brown D. A., Krishnan B., Reddy S. 2020, Nature Astronomy, 4, 625
\bibitem[\protect\citeauthoryear{Chen et al.}{2023}]{Chen+2023omegacen} Chen W., Freire P.~C.~C., Ridolfi A., Barr E.~D., Stappers B., Kramer M., Possenti A., et al., 2023, MNRAS, 520, 3847. doi:10.1093/mnras/stad029
\bibitem[\protect\citeauthoryear{Clark75}{1975}]{Clark75} Clark G. W. 1975, ApJL, 199, L143 
\bibitem[\protect\citeauthoryear{Cocozza et al.}{2006}]{Cocozza06} Cocozza G., Ferraro F., Possenti A., D'Amico N. 2006, ApJL, 641, L29
\bibitem[\protect\citeauthoryear{Colpi et al.}{2002}]{colpi02} Colpi M., Possenti A., Gualandris A. 2002, ApJL, 570, L85
\bibitem[\protect\citeauthoryear{Colpi et al.}{2003}]{colpi03} Colpi M., Mapelli M., Possenti A. 2003, ApJ, 599, 1260
\bibitem[\protect\citeauthoryear{Corongiu et al.}{2012}]{Corongiu12} Corongiu A., Burgay M., Possenti A., Camilo F., D'Amico N., Lyne A. G., Manchester R. N., et al., 2012, ApJ, 760, 100
\bibitem[\protect\citeauthoryear{Corongiu et al.}{2023}]{Corongiu23} Corongiu A., Venkatraman Krishnan V., Freire P. C. C., Kramer M., Possenti A., Geyer M., Ridolfi A., et al., 2023, A\&A, 671, A72
\bibitem[\protect\citeauthoryear{Correnti et al.}{2016}]{Correnti+2016} Correnti M., Gennaro M., Kalirai J.~S., Brown T.~M., Calamida A., 2016, ApJ, 823, 18. doi:10.3847/0004-637X/823/1/18
\bibitem[\protect\citeauthoryear{D'Amico et al.}{2002}]{damico02} D'Amico N., Possenti A., Fici L., Manchester R. N., Lyne A. G., Camilo  F., Sarkissian J. 2002, ApJL, 570, L89
\bibitem[\protect\citeauthoryear{Elson, Fall, \& Freeman}{1987}]{Elson+1987} Elson R.~A.~W., Fall S.~M., Freeman K.~C., 1987, ApJ, 323, 54. doi:10.1086/165807
\bibitem[\protect\citeauthoryear{Ferraro et al.}{2003}]{ferraro03} Ferraro F. R., Possenti A., Sabbi E., D'Amicho N. 2003, ApJ, 596, L211
\bibitem[\protect\citeauthoryear{Fregeau et al.}{2004}]{Fregeau+2004} Fregeau J.~M., Cheung P., Portegies Zwart S.~F., Rasio F.~A., 2004, MNRAS, 352, 1. doi:10.1111/j.1365-2966.2004.07914.x
\bibitem[\protect\citeauthoryear{Fregeau \& Rasio}{2007}]{Fregeau_Rasio_2007} Fregeau J.~M., Rasio F.~A., 2007, ApJ, 658, 1047. doi:10.1086/511809
\bibitem[\protect\citeauthoryear{Gratton et al.}{1997}]{Gratton+1997} Gratton R.~G., Fusi Pecci F., Carretta E., Clementini G., Corsi C.~E., Lattanzi M., 1997, ApJ, 491, 749. doi:10.1086/304987
\bibitem[\protect\citeauthoryear{Gratton et al.}{2003}]{Gratton+2003} Gratton R.~G., Bragaglia A., Carretta E., Clementini G., Desidera S., Grundahl F., Lucatello S., 2003, A\&A, 408, 529. doi:10.1051/0004-6361:20031003
\bibitem[\protect\citeauthoryear{Grondin et al.}{2023}]{Grondin+2023} Grondin S.~M., Webb J.~J., Leigh N.~W.C., Speagle J.~S., Khalifeh R.~J., 2023, MNRAS, 518, 3. doi:10.1093/mnras/stac3367
\bibitem[\protect\citeauthoryear{Harris}{1996, 2010 update}]{harris96}
  Harris, W. E. 1996, AJ, 112, 1487 (2010 update)
\bibitem[\protect\citeauthoryear{Heggie \& Hut}{2003}]{heggie03}
  Heggie D. C., Hut P. 2003, The Gravitational Million-Body Problem:
  A Multidisciplinary Approach to Star Cluster Dynamics (Cambridge:
  Cambridge University Press)
\bibitem[\protect\citeauthoryear{H{\'e}non}{1971a}]{henon1971a} H{\'e}non M., 1971, Ap\&SS, 13, 284. doi:10.1007/BF00649159
\bibitem[\protect\citeauthoryear{H{\'e}non}{1971b}]{henon1971b} H{\'e}non M.~H., 1971, Ap\&SS, 14, 151. doi:10.1007/BF00649201
\bibitem[\protect\citeauthoryear{Hobbs et al.}{2005}]{Hobbs+2005} Hobbs G., Lorimer D.~R., Lyne A.~G., Kramer M., 2005, MNRAS, 360, 974. doi:10.1111/j.1365-2966.2005.09087.x
\bibitem[\protect\citeauthoryear{Hurley, Pols, \& Tout}{2000}]{sse} Hurley J.~R., Pols O.~R., Tout C.~A., 2000, MNRAS, 315, 543. doi:10.1046/j.1365-8711.2000.03426.x
\bibitem[\protect\citeauthoryear{Hurley, Tout, \& Pols}{2002}]{bse} Hurley J.~R., Tout C.~A., Pols O.~R., 2002, MNRAS, 329, 897. doi:10.1046/j.1365-8711.2002.05038.x
\bibitem[\protect\citeauthoryear{Jonker et al.}{2004}]{jonker04} Jonker P. G., Galloway D. K., McClintock J. E., Buxton M., Garcia M., Murray S. 2004, MNRAS, 354, 666
\bibitem[\protect\citeauthoryear{Kiel et al.}{2008}]{Kiel+2008} Kiel P.~D., Hurley J.~R., Bailes M., Murray J.~R., 2008, MNRAS, 388, 393. doi:10.1111/j.1365-2966.2008.13402.x
\bibitem[\protect\citeauthoryear{King}{1966}]{King1966} King I.~R., 1966, AJ, 71, 64. doi:10.1086/109857
\bibitem[\protect\citeauthoryear{Kremer et al.}{2019}]{Kremer+2019} Kremer K., Chatterjee S., Ye C.~S., Rodriguez C.~L., Rasio F.~A., 2019, ApJ, 871, 38. doi:10.3847/1538-4357/aaf646
\bibitem[\protect\citeauthoryear{Kremer et al.}{2020}]{Kremer+2020bh} Kremer K., Ye C.~S., Chatterjee S., Rodriguez C.~L., Rasio F.~A., 2020, IAUS, 351, 357. doi:10.1017/S1743921319007269
\bibitem[\protect\citeauthoryear{Kremer et al.}{2020}]{Kremer+2020catalog} Kremer K., Ye C.~S., Rui N.~Z., Weatherford N.~C., Chatterjee S., Fragione G., Rodriguez C.~L., et al., 2020, ApJS, 247, 48. doi:10.3847/1538-4365/ab7919
\bibitem[\protect\citeauthoryear{Kremer et al.}{2021}]{Kremer+2021wd} Kremer K., Rui N.~Z., Weatherford N.~C., Chatterjee S., Fragione G., Rasio F.~A., Rodriguez C.~L., et al., 2021, ApJ, 917, 28. doi:10.3847/1538-4357/ac06d4
\bibitem[\protect\citeauthoryear{Kroupa}{2001}]{Kroupa2001} Kroupa P., 2001, MNRAS, 322, 231. doi:10.1046/j.1365-8711.2001.04022.x
\bibitem[\protect\citeauthoryear{Leigh \& Sills}{2011}]{leigh11} Leigh N., Sills A., 
MNRAS, 410, 2370
\bibitem[\protect\citeauthoryear{Leigh et al.}{2014}]{leigh14} Leigh N. W. C., L\:utzgendorf N., Geller A. M., Maccarone T. J., Heinke C., Sesana A. 2014, MNRAS, 444, 29
\bibitem[\protect\citeauthoryear{Leigh et al.}{2016}]{leigh16} Leigh N. W. C., L\:utzgendorf N., Geller A. M., Maccarone T. J., Heinke C., Sesana A. 2016, MNRAS, 444, 29
\bibitem[\protect\citeauthoryear{Leigh et al.}{2016b}]{leigh16b} Leigh N. W. C., Stone N. C., Geller A. M., Shara M. M., Muddu H., Solano-Oropeza D., Thomas Y. 2016b, MNRAS, 463, 3311
\bibitem[\protect\citeauthoryear{Leigh \& Wegsman}{2018}]{leigh18} Leigh N. W. C., Wegsman S. 2018, MNRAS, 476, 336
\bibitem[\protect\citeauthoryear{Leigh et al.}{2020}]{leigh20} Leigh N. W. C., Toonen S., Portegies Zwart S. F., Perna R. 2020, MNRAS, 496, 1819
\bibitem[\protect\citeauthoryear{Leigh et al.}{2022}]{leigh22} Leigh N. W. C., Stone N. C., Webb J.J., Lyra W. 2022, MNRAS, 517, 3838
\bibitem[\protect\citeauthoryear{Leitherer, Robert \& Drissen}{1992}]{leitherer92} Leitherer C., 
Robert C., Drissen L. 1992, ApJ, 401, 596
\bibitem[\protect\citeauthoryear{Lorimer}{2013}]{Lorimer13} Lorimer, D, 2013, Proceedings IAU, 291, 237, doi:10.1017/S1743921312023769 
\bibitem[\protect\citeauthoryear{Merritt}{2013}]{merritt13} Merritt D. 2013, Dynamics and Evolution of Galactic Nuclei (Princeton: Princeton University Press)
\bibitem[\protect\citeauthoryear{Pfahl et al.}{2002}]{Pfahl02} Pfahl E., Rappaport S., Podsiadlowski P. 2002, ApJ, 573, 283
%
\bibitem[\protect\citeauthoryear{Provencal et al.}{1998}]{provencal1998}
Provencal J.L., Shipman H. L., Hog E., Thejll P. 1998, ApJ, 494, 759


\bibitem[\protect\citeauthoryear{Rodriguez et al.}{2018a}]{Rodriguez+2018a} Rodriguez C.~L., Amaro-Seoane P., Chatterjee S., Rasio F.~A., 2018, PhRvL, 120, 151101. doi:10.1103/PhysRevLett.120.151101
\bibitem[\protect\citeauthoryear{Rodriguez et al.}{2018b}]{Rodriguez+2018b} Rodriguez C.~L., Amaro-Seoane P., Chatterjee S., Kremer K., Rasio F.~A., Samsing J., Ye C.~S., et al., 2018, PhRvD, 98, 123005. doi:10.1103/PhysRevD.98.123005
\bibitem[\protect\citeauthoryear{Rodriguez et al.}{2022}]{CMC1} Rodriguez C.~L., Weatherford N.~C., Coughlin S.~C., Amaro-Seoane P., Breivik K., Chatterjee S., Fragione G., et al., 2022, ApJS, 258, 22. doi:10.3847/1538-4365/ac2edf
\bibitem[\protect\citeauthoryear{Sigurdsson}{2003}]{sigurdsson03} Sigurdsson S. 2003, ASP Conference Proceedings, 302, 391
\bibitem[\protect\citeauthoryear{Sollima et al.}{2008}]{sollima08}
Sollima A., Beccari G., Ferraro F. R., Fusi Pecci F., Sarajedini
A. 2008, A\&A, 481, 701
\bibitem[\protect\citeauthoryear{Souza et al.}{2020}]{Souza+2020} Souza S.~O., Kerber L.~O., Barbuy B., P{\'e}rez-Villegas A., Oliveira R.~A.~P., Nardiello D., 2020, ApJ, 890, 38. doi:10.3847/1538-4357/ab6a0f
\bibitem[\protect\citeauthoryear{Tauris \& Savonije}{1999}]{TaurisSavonije99} Tauris T. M., Savonije G.J., 1999, A\&A, 350, 928
\bibitem[\protect\citeauthoryear{Tremblay et al.}{2016}]{tremblay2016} Tremblay P.-E., Cummings J., Kalirai J., Gansicke B.-T., Gentile-Fusillo N., Raddi R. 2016, MNRAS, 461, 2100

\bibitem[\protect\citeauthoryear{Tokovinin}{2018}]{tokovinin18} Tokovinin A. 2018, ApJS, 235, 6
\bibitem[Valtonen \& Karttunen(2006)]{valtonen06} Valtonen, M., \& Karttunen, H.\ 2006, The Three-Body Problem, by Mauri Valtonen and Hannu Karttunen, pp.~.~ISBN 0521852242.~Cambridge, UK: Cambridge University Press,  2006
\bibitem[\protect\citeauthoryear{Verbunt \& Freire}{2014}]{verbunt_freire_2014} Verbunt F., Freire P.~C.~C., 2014, A\&A, 561, A11. doi:10.1051/0004-6361/201321177
\bibitem[\protect\citeauthoryear{Wang et al.}{2019}]{wang18} Wang Y.-H., Leigh N., Sesana A., Perna R. 2019, MNRAS, 482, 3206
\bibitem[\protect\citeauthoryear{Webb et al.}{2018}]{webb18} Webb J. J., Leigh N. W. C., Singh A., Ford K. E. S., McKernan B., Bellovary J. 2018, MNRAS, 474, 3835
\bibitem[\protect\citeauthoryear{Webb et al.}{2019}]{webb19} Webb J. J., Leigh N. W. C., Serrano R., Bellovary J., Ford K. E. S., McKernan B., Spera M., Trani A. A. 2019, MNRAS, 488, 3055
\bibitem[\protect\citeauthoryear{Ye et al.}{2019}]{Ye+2019} Ye C.~S., Kremer K., Chatterjee S., Rodriguez C.~L., Rasio F.~A., 2019, ApJ, 877, 122. doi:10.3847/1538-4357/ab1b21
\bibitem[\protect\citeauthoryear{Ye et al.}{2022}]{Ye+2022} Ye C.~S., Kremer K., Rodriguez C.~L., Rui N.~Z., Weatherford N.~C., Chatterjee S., Fragione G., et al., 2022, ApJ, 931, 84. doi:10.3847/1538-4357/ac5b0b
\bibitem[\protect\citeauthoryear{Ye et al.}{2023}]{Ye+2023} Ye C.~S., Kremer K., Ransom S.~M., Rasio F.~A., 2023, arXiv, arXiv:2307.15740. doi:10.48550/arXiv.2307.15740
\bibitem[\protect\citeauthoryear{Zhao \& Heinke}{2022}]{Zhao22} Zhao J., Heinke C. O., 2022, MNRAS, 511, 5964, doi:10.1093/mnras/stac442

\end{thebibliography}
\end{document}